\documentclass[12pt]{article}

\usepackage{amsmath,amssymb}
\usepackage{graphicx}

\makeatletter
 
  \@addtoreset{equation}{section}
\makeatother

\newcommand{\be}{\begin{equation}}
\newcommand{\ee}{\end{equation}}
\newcommand{\bea}{\begin{eqnarray}}
\newcommand{\eea}{\end{eqnarray}}
\newcommand{\beann}{\begin{eqnarray*}}
\newcommand{\eeann}{\end{eqnarray*}}
\newcommand{\nn}{\nonumber}
\newcommand{\ba}{\begin{array}}
\newcommand{\ea}{\end{array}}

\newcommand{\Tr}{{\rm Tr}\,}
\newcommand{\del}{\partial}
\newcommand{\Z}{\mathbb{Z}}
\newcommand{\R}{\mathbb{R}}
\newcommand{\C}{\mathbb{C}}

\newcommand{\IN}{\mathbb{N}}
\newcommand{\kv}{\mbox{\boldmath$k$}}
\newcommand{\av}{\mbox{\boldmath$a$}}
\newcommand{\mv}{\mbox{\boldmath$m$}}
\newcommand{\e}{\epsilon}
\newcommand{\CP}{\mathbb{C}{\rm P}}

\newcommand{\N}{{\cal N}}
\newcommand{\D}{{\cal D}}
\newcommand{\pint}{-\hspace{-13pt}\int}
\newcommand{\drho}{{\mit\Delta}\rho}

\newcommand{\cN}{{\cal N}}
\newcommand{\gYM}{{{\rm gYM}_2}}

\newcommand{\tk}{\widetilde{k}}
\newcommand{\cO}{{\cal O}}

\newcommand{\cD}{{\cal D}}

\newcommand{\ta}{\widetilde{a}}

\newcommand{\tB}{{\widetilde{B}}}

\newcommand{\tg}{{\widetilde{g}}}

\begin{document}

\begin{titlepage}

\setcounter{page}{0}
\renewcommand{\thefootnote}{\fnsymbol{footnote}}

\begin{flushright}
hep-th/0504176 \\
RIKEN-TH-40
\end{flushright}

\vspace{15mm}
\begin{center}
{\large\bf
 Localization on the D-brane, two-dimensional gauge theory and matrix models
 }

\vspace{18mm}
{
So Matsuura\footnote{{\tt matsuso@riken.jp}} and
Kazutoshi Ohta\footnote{{\tt k-ohta@riken.jp}}}\\
\vspace{10mm}

{\em Theoretical Physics Laboratory\\
The Institute of Physical and Chemical Research (RIKEN)\\
2-1 Hirosawa, Wako\\
Saitama 351-0198, JAPAN}\\

\end{center}

\vspace{18mm}
\centerline{{\bf Abstract}}
\vspace{10mm}
\small{%
We consider the effective topological field theory on
Euclidean D-strings wrapping on a 2-cycle in the internal space.
We evaluate the vev of a suitable operator 
corresponding to the chemical potential of vortices
bounded to the D-strings,
and find that it reduces to the partition function
of generalized two-dimensional Yang-Mills theory 
as a result of localization.
We argue that the partition function gives a grand canonical ensemble
of multi-instanton corrections for four-dimensional $\N{=}2$ gauge theory
in a suitable large $N$ limit.
We find two-dimensional gauge theories that provide the instanton
partition function
for four-dimensional $\N{=}2$ theories with the hypermultiplets in
the adjoint and fundamental representations.  
We also propose a partition function that gives
the instanton contributions to four-dimensional $\N{=}2$ quiver gauge
theory.
We discuss the relation between
Nekrasov's instanton partition function and the Dijkgraaf-Vafa theory 
in terms of large $N$ phase transitions of
the generalized two-dimensional Yang-Mills theory.}

\end{titlepage}
\newpage

\renewcommand{\thefootnote}{\arabic{footnote}}
\setcounter{footnote}{0}

\section{Introduction and Summary}

One of the most significant developments 
in supersymmetric (SUSY) gauge theories in recent years is that
non-perturbative dynamics including instanton corrections becomes 
to be captured by interesting physical/mathematical systems.
In particular, the prepotential of four-dimensional 
$\N{=}2$ SUSY gauge theory 
and the superpotential of $\N{=}1$ SUSY gauge theory, 
which are both holomorphic,
are obtained from the free energy of a random partition
\cite{Nekrasov:2003rj}
and a random matrix model \cite{Dijkgraaf:2002fc, Dijkgraaf:2002vw,
Dijkgraaf:2002dh},
respectively. 
It is also pointed out that the random partition describing the
prepotential of the $\N{=}2$ theory is related to 
a large $N$ limit of two-dimensional Yang-Mills theory \cite{Matsuo:2004cq}.
Similar developments have also come to topological 
string theories. 
For example, amplitudes on a singular Calabi-Yau manifold 
are obtained from some statistical model like the melting crystal
\cite{Okounkov:2003sp, Iqbal:2003ds}. 
Furthermore, the closed topological string amplitude also has 
a dual open string picture which is described by 
the large $N$ limit of three-dimensional Chern-Simons gauge theory
\cite{Gopakumar:1998vy, Gopakumar:1998ki}.

An interesting fact is that these systems  
are ``bosonic'' even though
the original gauge theories and the topological string theories 
are preserving SUSY. 
In prior to the discovery of the relationship between the holomorphic
quantities of the SUSY systems and the bosonic partition functions,
relations between SUSY 
gauge theories and ``bosonic'' integrable systems are already pointed out
(see for review \cite{Gorsky:2000px}). 
In fact, 
the Seiberg-Witten curve of the four-dimensional $\N{=}2$ SUSY gauge theory
is especially the spectrum curve for the integrable system
and vacua in the four-dimensional $\N{=}1$ theory are
equilibrium points of the related
integrable system \cite{Hollowood:2003ds}.

The key to understand these relations should exist 
in a localization of the path integral 
in an effective theory on D-branes. 
The non-perturbative corrections to the SUSY gauge theory
are coming from corrections of Euclidean BPS D-branes, 
whose world-volume does not contain the time
direction and which are wrapping on non-trivial cycles
in an internal space, 
if we embed the gauge theory into a string
theoretical configuration with a suitable compactification.
According to \cite{Bershadsky:1995qy}, the effective theory on the
Euclidean D-brane must be topologically twisted.
Moreover, if we use the feature of the topological field theory, 
the partition function is reduced to a ``bosonic'' system after
integrating out all fermionic fields due to the localization
\cite{Witten:1992xu, Moore:1997dj}.

In this paper, we discuss the relation between SUSY gauge theories 
and two-dimensional (bosonic) gauge theories from the first principle, that is,  
from the view point of the localization. 
We first give a configuration of D-branes that realize 
an $\N{=}2$ SUSY gauge theory and
consider the effective theory on the D-branes wrapping on
the internal 2-cycle in section 2. 
We evaluate the expectation value of an operator, 
which corresponds to the chemical potential 
of vortices on the cycle, 
and show that it reduces to the BF-theory. 
If we deform generically the D-brane effective theory 
by adding a ``potential'' term,
we obtain the partition function of
the so-called generalized two-dimensional Yang-Mills ($\gYM$) theory
\cite{Ganor:1995bq, Dijkgraaf:2003xk}.
This reduction is essential reason why the prepotential of the SUSY 
gauge theory and the partition function of the two-dimensional
Yang-Mills theory  
are related with each other. 
In this section, we also discuss the relation between our analysis 
and the argument given in \cite{Vafa:2004qa, Aganagic:2004js} 
where the authors derive the $q$-deformed two-dimensional Yang-Mills theory 
using a similar setup. 

Moreover, in section 3, we proceed the path integral of $\gYM$ following
the standard gauge theoretical method \cite{Blau:1993tv, Blau:1993hj}.
We will see that the system finally reduces to a ``discrete matrix
model''. The difference between the discrete matrix model 
and an ordinary 
random matrix model, which is a model for $c{=}0$ string theory and
effective superpotential calculations by Dijkgraaf and Vafa, is that
eigenvalues of the discrete matrix model are discretized in a unit while
the eigenvalues of the ordinary matrix model are continuous variables.  Each
eigenvalue can not exist at the same position (value) due to the
existence of the Vandermonde determinant. So a possible set of
eigenvalues is given by a strongly decreasing (non-colliding) integer 
sequence.  Thus the integral over possible eigenvalues in the partition
function of the random matrix model is replaced by a summation over
possible sets of non-colliding ordered integers.  On the other hand, if
we map these discrete eigenvalue distributions to a weakly decreasing
sequence by a suitable shift, we can identify the integer sequence with
numbers of rows of a Young diagram. 
Using this identification, 
we can see the relation between the discrete matrix
model and Migdal's partition function \cite{Migdal:1975zg}, 
which is expressed as a summation over sets of 
the Young diagrams (representations of $U(N)$). 
In addition, we regard the eigenvalues as states of free
fermions obeying the exclusion principle by using the famous Young diagram/Maya
diagram (free fermion states) correspondence \cite{Maeda:2004is}.

We find that the discrete matrix model possesses an essential structure
of Nekrasov's instanton partition function. 
If we consider a potential term, 
the eigenvalues of the discrete matrix model are accumulated around the
critical points of the potential, then  
each lump of the eigenvalues has two end points (fermi surfaces). 
We show that a suitable large $N$ limit 
decouples these fermi surfaces with each other. 
Then the partition function is factorized into two sectors, 
and one of them produces the instanton partition 
function whose rank of the gauge group and moduli parameters of $\N{=}2$
theory are determined by the critical points of the $\gYM$ potential. 
This manner to take the large $N$ limit is similar to the chiral
decomposition by Gross and Taylor
\cite{Gross:1993hu,Gross:1993yt,Gross:1993cw} (and see for review
\cite{Cordes:1995fc}).

In section 4, we extend the arguments in section 3 
to the $\cN{=}2$ theories including the hypermultiplets in
various representations. 
We first derive Nekrasov's partition function of theory 
with the hypermultiplet in the adjoint representation 
given in \cite{Bruzzo:2002xf,Nekrasov:2003rj}
from a suitable two-dimensional model.  
To obtain the initial two-dimensional theory, 
we add extra observables to the topological theory, 
and the additional terms have the same form as the (tree level) 
superpotential of the corresponding four-dimensional theory. 
This extension very looks like the Dijkgraaf-Vafa's construction
to determine the matrix model actions, which is 
simply obtained from the tree level superpotential by replacing the
superfields with the hermitian matrices. 
Then we get Nekrasov's partition function of this model
from the large $N$ limit again. 
Once we obtain the manner to make the observables for adjoint matters, 
it is easy to generalize it to the quiver theory%
\footnote{The multi-instanton calculus for the quiver gauge
theory is proposed by \cite{Fucito:2004gi} 
from a point of view of the equivariant cohomology.}
and the theory with the hypermultiplets 
in the fundamental representation, 
since these theories are essentially reproduced by flows 
from the theory with the adjoint matter. 
We exhibit an $U(r)\times U(s)$ $A_2$-type quiver theory 
and an $U(r)$ gauge theory with $s$ flavors in
the fundamental representation.

In section 5, we discuss the relationship between the discrete matrix
model and usual continuous matrix model more explicitly. There we see
the discrete matrix model in the continuum and large $N$ limit 
has two different type of third order phase transitions. 
One is called as the Douglas-Kazakov phase transition
\cite{Douglas:1993ii} and another is
the Gross-Witten phase transition \cite{Gross:1980he}. We give a
correspondence between the limiting shape of the Young diagram and
the eigenvalue density (free fermion states). We find that the
Douglas-Kazakov and Gross-Witten phase transition are dual under
electron/hall exchanges (exchanges of the eigenvalue and vacant
positions). Combining these observations with the arguments in section 3,
we also find the continuum and large $N$ limit to derive the Nekrasov's
instanton partition function and Dijkgraaf-Vafa's matrix model analysis
should lie on different phases of the discrete matrix model. We expect
that relations among various phases may connect non-perturbative dynamics
of $\N{=}2$ and $\N{=}1$ theories.

\section{Localization on D-brane and two-dimensional YM theory}
\label{sec:localization-d-brane}

Four-dimensional gauge theories with eight supercharges can be realized
in string theory in various ways like the geometric engineering
\cite{Katz:1996fh,Katz:1997eq}
and
Hanany-Witten type brane configuration \cite{Hanany:1997ie,Witten:1997sc}. 
Among them, we consider Type IIB superstring theory on $\R^{1,3}\times
\C \times {\cal M}_4$, where ${\cal M}_4$ is a four-dimensional ALE
space.
If
D5-branes are wrapped on 2-cycles in ${\cal M}_4$, a four-dimensional
gauge theory will appear on the extra $\R^{1,3}$ space on the D5-branes
except for the compactified internal 2-cycles.  
The configuration for the four-dimensional
gauge theory preserves 8 supercharges at least when the 2-cycle is
$\CP^1$, which is T-dual to the
Hanany-Witten Type IIA brane configuration of
the four-dimensional $\N{=}2$ SUSY gauge
theory. So it is sufficient to consider the case of $\CP^1$ from the
field theoretical point of view, but we
will treat genus of the 2-cycle as generic one throughout the paper.
So we assume the structure near a single cycle is $T^*\Sigma_G$,
where $\Sigma_G$ is a Riemann surface with genus $G$, 
that is, 
the world-volume of the D5-branes
is thought to be spanned along $\R^{1,3}\times \Sigma_G$.
The gauge coupling of the four-dimensional system relates to the area $A$
of $\Sigma_G$ by
\be
\frac{1}{g_{\rm YM_4}^2} = \frac{A}{g_s l_s^2}.
\ee
We set $l_s=1$ hereafter. 

In this article, we are interested in contribution of instantons  
to the partition function of the four-dimensional $\cN{=}2$ 
SUSY gauge theory. 
To this end, we start from the worldvolume theory of D5-branes on
$\R^{1,3}\times \Sigma_G$ described above.  
It is known that the theory admits noncommutative deformation on
$\R^{1,3}$ and it does not affect to the final result of the
prepotential \cite{Nekrasov:1998ss,Nekrasov:2002qd}.
According to the concept of the large $N$ reduction due to the
noncommutativity
\cite{Eguchi:1982nm,Ishibashi:1997xs,Seiberg:2000zk}, 
the worldvolume theory on D5-branes is reduced to the two-dimensional
gauge theory on a large number of Euclidean D-strings wrapping on
$\Sigma_G$. 
As discussed in \cite{Bershadsky:1995qy}, 
the low energy effective theory on D-branes wrapping on the cycle 
must be (partially) topologically twisted 
to preserve the SUSY. 
In this case, the topological theory should be localized on 
the Hitchin system, 
\begin{align}
 F_{z\bar{z}} = \left[W_z,\overline{W}_{\bar{z}}\right],\qquad 
 \overline{D}_{\bar{z}} W_{z} = D_z \overline{W}_{\bar{z}}=0\,, 
\end{align}
where $W_z$ and $\overline{W}_{\bar{z}}$ are 1-forms on $\Sigma_G$
associated with the normal bundle of $\Sigma_G$. 
Notice that these fields correspond to the fluctuation 
of $\Sigma_G$ in ${\cal M}_4$.
However, we can simplify the analysis by giving a huge mass
to $W_z$ \cite{Vafa:2004qa}. 
Correspondingly, we freeze the degrees of freedom as  
$W_z=\overline{W}_{\bar{z}}=0$ and 
the system is localized on the flat connection $F=0$.

The twisted theory on the internal cycles enjoys the BRST symmetry
and contains essentially a two-dimensional gauge field
$A\equiv A_idx^i$ ($i=1,2$),
its (fermionic) BRST partner
$\lambda\equiv \lambda_idx^i$
and a complex scalar field $\Phi$ in the adjoint representation of
the gauge group $U(N)$ which 
corresponds to the position of D-strings in  
the flat directions $\C$.  
These fields transform under the BRST symmetry as%
\footnote{
$Q$ acts on $X$ as a commutator or anti-commutator
depending on whether $X$ is bosonic or fermionic.
} 
\be
Q A = \lambda, \quad Q \lambda = -d_{A} \Phi,
\quad Q \Phi = 0,
\label{BRST1}
\ee
where $d_{A}\Phi\equiv d\Phi + \left[A,\Phi\right]$. 
The BRST operator $Q$ is nilpotent up to the gauge transformation,
and $A$, $\lambda$ and $\Phi$ have the ghost (BRST) charges of
0,1 and 2, respectively.
As mentioned above, we need to impose the
condition $F\equiv dA + A\wedge A=0$. 
To make this constraint and write
down Lagrangian, we need additional BRST multiplets $(\bar{\Phi},\eta)$
and $(H,\chi)$, which transform as 
\be
\begin{array}{ll}
Q \bar{\Phi} = \eta, & Q \eta = [\Phi,\bar{\Phi}],\\
Q H = [\Phi,\chi], & Q \chi = H.
\end{array}
\label{BRST2}
\ee
Using these BRST multiplets, 
the action of the twisted theory can be written
in the BRST exact form as 
\be
S_{\rm top}=\frac{1}{h^2}Q\int_{\Sigma_G} 
\Tr\Xi(A,\lambda,\Phi,\bar{\Phi},\eta,H,\chi), 
\label{topological action}
\ee
where $h^2$ is a coupling constant which is proportional to 
the string coupling $g_s$ and 
\be
\Xi
= \frac{1}{4}\eta[\Phi,\bar{\Phi}]
+\frac{1}{2}\chi(H-2\ast F)
+d_A\bar{\Phi}\wedge \lambda. 
\label{Xi}
\ee
If we write the bosonic and fermionic fields together as
$\vec{\cal B}=(A,H,\bar{\Phi})$ and
$\vec{\cal F}=(\lambda,\chi,\eta)$, respectively
(note that $\Phi$ does not have any fermionic partner),
the partition function is given by a path integral
over these fields;
\be
Z_{\rm top} = \int \D \vec{\cal B} \D \vec{\cal F} \D \Phi
e^{-\frac{1}{h^2}Q\int_{\Sigma_G} 
\Tr\Xi(\vec{\cal B}, \vec{\cal F},\Phi)
}.
\label{top path int}
\ee
Here we notice that the action (\ref{topological action}) can be obtained 
from the dimensional or large $N$ reduction from the four-dimensional 
twisted $\N{=}2$ gauge theory,
where an adjoint scalar field and a Lagrange
multiplier corresponding to the moduli and constraints of the normal
direction to the 2-cycle are thrown away by turning on a huge mass
\cite{Witten:1988ze}.

From the variation of the partition function with respect to
the coupling constant $h^2$, we find the partition function 
(\ref{top path int}) is independent of the coupling. 
This means that the path integral
can be evaluated exactly in the WKB (weak coupling) limit.
So the path integral dominates around the Gaussian integrals with the
constraints.
In particular, by integrating out $H$, this integral gives
a flat connection constraint $F=0$. 
Therefore, 
any physical observables
must be evaluated around the flat connections 
and the path integral (\ref{top path int}) 
gives schematically the ``volume'' of the moduli 
space of the flat connections, namely the genus $G$ Riemann surface
with marked points.

Next we consider BPS bound states on the D-strings.
In the context of the topological field theory, 
we can introduce the chemical potentials 
for the BPS objects bounded on the D-strings
by evaluating the vacuum expectation value
of observables in the topological field theory \cite{Vafa:2004qa}. 
In general, observables in topological field theory
are cohomology classes of the BRST operator, 
which are constructed from $p$-forms ${\cal O}_p$ obeying 
the descent equations \cite{Witten:1988ze}, 
\be
\begin{array}{l}
d_A{\cal O}_0 = Q{\cal O}_1, \qquad 
d_A{\cal O}_1 = Q{\cal O}_2, 
\end{array}
\ee
where, 
\be
\begin{array}{l}
{\cal O}_0 = \frac{1}{2}\Tr \Phi^2,\\
{\cal O}_1 = \Tr \Phi \lambda,\\
{\cal O}_2 = \Tr \left(
i\Phi F +\frac{1}{2}\lambda \wedge \lambda
\right).
\end{array}
\ee 
Let us first consider the integral of 2-form ${\cal O}_2$;
\be
I_2(\Sigma_G) = \int_{\Sigma_G}
\Tr \left(i\Phi F +\frac{1}{2}\lambda \wedge \lambda
\right), 
\ee
and consider the vacuum expectation value, 
\be
\left\langle
e^{
-\frac{1}{g_s}I_2(\Sigma_G)
}
\right\rangle_{\rm top}
\equiv
\int \D \vec{\cal B} \D \vec{\cal F} \D \Phi
\exp\left\{
\int_{\Sigma_G}
\Tr
\left[
-\frac{1}{h^2}Q\Xi  
-\frac{1}{g_s}\left(
i\Phi F + \lambda\wedge\lambda
\right)
\right]
\right\}, 
\label{chemical potential}
\ee 
in the topological field theory discussed above.
This observable corresponds to the chemical potential of
vortices on the D-strings.%
\footnote{
In the context of the theory of D4-branes considered in 
\cite{Vafa:2004qa}, $I_2(\Sigma_G)$ relates to the chemical potential
of D0-branes by T-duality.
Hence the vortices corresponds to the intersection of D-strings.
The authors thank to the referee of Physical Review D
for pointing it out. 
}
In fact, in the next section,
we will show that $I_2(\Sigma_G)$ introduces 
a coupling of the eigenvalues of $\Phi$
with the instanton numbers corresponding to the maximally
broken gauge symmetry $U(1)^N$. 
The localization onto the moduli space of the flat connection
due to the topological action does not affect to the physical
part of the action
$\frac{1}{g_s}\int_{\Sigma_G}\left(i \Phi F + \lambda\wedge\lambda\right)$,
since the e.o.m. of the BF-theory also gives the flat connection
constraint $F=0$.
So, after integrating out the bilinear of $\lambda$, we find
(\ref{chemical potential}) exactly reduces to
a partition function of the bosonic BF theory \cite{Witten:1992xu}, 
\be
\left\langle
e^{
-\frac{1}{g_s}I_2(\Sigma_G)
}
\right\rangle_{\rm top}
=\int{\cal D}A{\cal D}\Phi
\exp\left[
-\frac{1}{g_s}\int_{\Sigma_G}
\Tr i\Phi F
\right].
\ee

We can also consider zero-form observables in the topological field. 
Especially, we consider operators 
in the form of the trace of polynomials of $\Phi$, 
which are actuarially observables since $\Phi$ is BRST closed itself. 
Recall that $\Phi$ corresponds to the positions of D5-branes 
in the $\C$-direction, which are the moduli parameters 
of vacua in Coulomb phase of the four-dimensional theory. 
To fix the moduli parameters in the four-dimensional theory, 
we add a generic $(r+1)$-the order superpotential 
for the adjoint scalar field manually,
then the moduli parameters are expected to be fixed 
around the critical points of the superpotential 
\cite{Dijkgraaf:2002fc}.%
\footnote{
In order to recover the $\cN{=}2$ supersymmetry, 
we reduce the superpotential to zero adiabatically
after fixing the moduli parameters.}
We can realize this procedure in the two-dimensional theory
by deforming the expectation value (\ref{chemical potential})
by the observable, 
\be
\Tr W(\Phi)\equiv \sum_{j=1}^{r+1}\frac{\mu_j}{j}\Tr \Phi^j. 
\ee
In contrast to the previous case without potential,
we cannot claim that
the expectation value of the observable deformed by the potential, 
$I_2(\Sigma_G)+\int_{\Sigma_G}\Tr W(\Phi)\omega$ 
(where $\omega$ is a volume form on $\Sigma_G$),  
is the same as the partition function of the deformed BF theory, 
since higher critical points corresponding to $F\neq 0$ may contribute
to the path integral. However, as discussed in \cite{Witten:1992xu}, the
contributions from the higher critical points are exponentially
small. 
Thus, after integrating out $\lambda$,
we can again evaluate approximately the expectation value in 
the topological field theory as the partition function of 
a ``physical'' theory up to the contribution from the higher critical
points;
\begin{multline}
\left\langle
\exp\left[
-\frac{1}{g_s}\int_{\Sigma_G} \Tr\left(
i\Phi F + \lambda\wedge\lambda 
\right)
-\frac{1}{g_s}\int_{\Sigma_G} \Tr W(\Phi) \omega
\right]
\right\rangle_{\rm top}\\
\cong \int{\cal D}A {\cal D}\Phi
\exp\left[
-\frac{1}{g_s}\int_{\Sigma_G}
\Tr \Bigl(
i\Phi F
+W(\Phi)\omega
\Bigr)
\right].
\end{multline}
Fortunately, however,
the e.o.m of the r.h.s. theory with respect to $\Phi$ gives  
\be
iF = W'(\Phi)\omega. 
\ee
So we can expect that the above approximation becomes much better,
if we take into account configurations only 
around the critical points $W'=0$, where the contribution from the flat
connection $F=0$ is dominated.

Hereafter we investigate the bosonic BF type theory deformed by the
potential as an effective theory of the D-strings with vortices. 
This model is known as the generalized two-dimensional Yang-Mills
theory ($\gYM$)
\cite{Ganor:1995bq},
 which is a generalization of the ordinary two-dimensional $U(N)$
Yang-Mills theory.
Indeed, if we choose a quadratic potential
$W(\Phi)=\frac{\mu_2}{2}\Phi^2$, 
the above partition function reduces to the usual
two-dimensional Yang-Mills theory after integrating out $\Phi$,
\be
Z_{\rm YM_2}=\int \D A
\exp\left[
-\frac{1}{2g_{\rm YM_2}^2} \int d^2x \Tr F^2
\right],
\ee
with an identification of $g_{\rm YM_2}=g_s \mu_2$.

Before closing this section, we need to mention on the relationship
between our model and a system argued in \cite{Aganagic:2004js}. 
In the context of \cite{Aganagic:2004js}, 
the adjoint scalar $\Phi$ 
comes from the holonomy of the gauge field at infinity 
of a fiber direction, 
which causes the periodicity of the field $\Phi \sim \Phi + 2\pi$. 
Then one needs to use a unitary measure for it and 
the model gets the $q$-deformation.
On the other hand, in our model, we do not assume 
such a periodicity in the $\C$-direction, 
which is associated with the vev for
the adjoint scalar field $\Phi$. 
Namely, a range of value of $\Phi$ is non-compact and we do not need to
use the unitary measure. 
In other words, 
our model is thought to be a decompacified limit of the periodic
direction in the fiber. 
So we have the $\gYM$, not the $q$-deformed one.

\section{Instanton counting from two-dimensional gauge theory}
\label{sec:inst-count-from}

\subsection{Migdal's partition function (Gaussian model)}
\label{sec:migd-part-funct}

In prior to investigating the partition function of 
the $\gYM$,
we derive Migdal's partition function \cite{Migdal:1975zg}
for the ordinary two-dimensional Yang-Mills theory
by using the Abelianization technique developed in
\cite{Blau:1993tv,Blau:1993hj}. 
As mentioned in the previous section, 
the partition function of the two-dimensional Yang-Mills theory 
can be written as the $\gYM$ with the quadratic potential, 
\be
Z=\int {\cal D}A {\cal D}\Phi
\exp\left[
-\frac{1}{g_s}\int_{\Sigma_G}d^2z
\Tr \left(
i\Phi F + \frac{\mu}{2}\Phi^2
\right)
\right]. 
\ee
We now decompose the Lie algebra valued fields $\Phi$ and $A$ into
\be
\Phi = \sum_i \Phi_i T_i
+ \sum_\alpha \Phi_\alpha E_\alpha,
\qquad
A = \sum_i A_i T_i
+ \sum_\alpha A_\alpha E_\alpha,
\ee
where $T_i$'s and $E_\alpha$'s are 
the Cartan subalgebra and the root, respectively. 
We introduce the following Faddeev-Popov determinant 
corresponding to the gauge fixing $\Phi_\alpha=0$;  
\bea
\Delta_{\rm FP}^{-1} &\equiv& 
\int {\cal D}\theta
\prod_{x,\alpha}\delta(\Phi^\theta_\alpha(z))\nn\\
&=& {\rm Det}^{-1}\,(i \sum_i \alpha_i\Phi_i(z)),
\eea
where $\alpha_i$ is a root vector 
determined by $[T_i,E_\alpha]=\alpha_iE_\alpha$.
Then we get
\be
Z=\int
{\cal D}A_i{\cal D}A_\alpha{\cal D}\Phi_i
{\cal D}c_\alpha{\cal D}\bar{c}_{\alpha}
e^{-\frac{1}{g_s}(S_{\rm GF} + S_{\rm ghost})},
\label{ghost int}
\ee
where
\bea
S_{\rm GF} &=&
\int_{\Sigma_G}\sum_i\left(
i\Phi_i dA_i 
+ i\sum_\alpha(\alpha\cdot\Phi)A_\alpha\wedge A_{-\alpha}
+ \frac{\mu}{2}\Phi_i^2 \omega
\right), \\
S_{\rm ghost} &=& 
\int_{\Sigma_G}i\sum_\alpha(\alpha\cdot\Phi)
\bar{c}_{-\alpha} c_{\alpha}\omega.
\eea

We first integrate out one forms $A_\alpha$
and complex scalars $(\bar{c}_{-\alpha},c_\alpha)$.
According to the Hodge decomposition theorem,
any $p$-form $\xi_p$ on a compact orientable
Riemann surface $\Sigma_{G}$ can be uniquely decomposed as 
\be
\xi_p = d\alpha_{p-1} + d^\dag\beta_{p+1} + \gamma,
\ee
where $d^\dag=-*d*$, $\alpha_{p-1}\in\Omega^{p-1}(\Sigma_G)$,
$\beta_{p+1}\in\Omega^{p+1}(\Sigma_G)$
and $\gamma$ is a harmonic $p$-form.
Applying this to $A_\alpha$
and $(\bar{c}_{-\alpha},c_\alpha)$, we find that
the number of modes of $A_\alpha$ and $(\bar{c}_{-\alpha},c_\alpha)$
is $2\dim \Omega^0(\Sigma_G)+\dim H^1(\Sigma_G)$
and  $2\dim \Omega^0(\Sigma_G)+2\dim H^0(\Sigma_G)$, respectively.
So, all non-harmonic
modes cancel each other out and only harmonic zero modes,
whose number is equal to
$2\dim H^0(\Sigma_G)-\dim H^1(\Sigma_G)=\chi(\Sigma_G)$, contribute
to the path integral in (\ref{ghost int}).
So we obtain the Abelian gauge theory with the partition function, 
\be
Z=\int \prod_{i}{\cal D}A_i{\cal D}\Phi_i
\prod_{i\neq j}(\Phi_i-\Phi_j)^{1-G}
\exp\left[
-\frac{1}{g_s}\int_{\Sigma_G}
\sum_i\left(
i \Phi_i dA_i +\frac{\mu}{2}\Phi_i^2\omega
\right)
\right],
\ee
where we use $\chi(\Sigma_G)=2-2G$.

Recall that the BRST transformation of $\Phi$ in (\ref{BRST1}),
the localization at fixed points of the BRST transformations, 
or the Gauss law constraint, 
tells us that the diagonal $U(1)$ parts $\Phi_i(z)$ are holomorphic
everywhere on the closed Riemann surface $\Sigma_G$.
This means that net effect of the path integral over $\Phi_i$
is coming from an independent part in $z$.
Therefore,
we can put $\Phi_i(z)=\lambda_i$, where $\lambda_i$ are $z$-independent
variables. So the partition function becomes
\be
Z=\int \prod_{i}{\cal D}A_i d\lambda_i
\prod_{i\neq j}(\lambda_i-\lambda_j)^{1-G}
\exp\left[
-\frac{1}{g_s}\int_{\Sigma_G}
\sum_i\left(
i \lambda_i dA_i +\frac{\mu}{2}\lambda_i^2\omega
\right)
\right].
\ee
Thus the integration over the gauge fields becomes
simply a summation over $U(1)^N$ non-trivial maximal torus bundles,
which are classified by the first Chern class, 
\be
\int_{\Sigma_G} dA_i = 2\pi p_i,  \quad p_i\in \Z.
\ee
Then the combination $\frac{\lambda_i}{2\pi g_s}$ plays a role of
the chemical potential for vortices on $i$th D-string 
wrapping on the 2-cycle  
as mentioned in the previous section. 
The integration over $A_i$ reduces to the
summation over a set of integers $p_i$ $(i=1,\ldots,N)$ as 
\begin{align}
Z&=\frac{1}{N!}\prod_{k}\sum_{p_k\in \Z}\int d\lambda_k
\prod_{i < j}(\lambda_i-\lambda_j)^{2-2G}
\exp\left[
-\frac{1}{g_s}
\left(
2 \pi i \lambda_k p_k +\frac{\mu A}{2}\lambda_k^2
\right)
\right] \nn \\
&=\frac{1}{N!}\prod_{k}\int d\lambda_k
\prod_{i < j}(\lambda_i-\lambda_j)^{2-2G}
\sum_{n_k\in\Z}\delta(\lambda_k/g_s-n_k)
\exp\left[
-\frac{\mu A}{2 g_s}\lambda_i^2
\right] \nn\\
&=\frac{1}{N!}\sum_{n_k\in\Z}
\prod_{i < j}(g_s n_i-g_s n_j)^{2-2G}
\exp\left[
-\frac{g_s \mu A}{2}\sum_i n_i^2
\right].
\end{align}
where the factor $1/N!$ corresponds to the Weyl denominator 
in an identification of $U(N)\simeq U(1)\times SU(N) /\Z_N$, 
and we have used the Poisson resummation formula, 
\be
\sum_{p_i\in\Z}
\exp\left(
-2\pi i\frac{\lambda_i}{g_s}p_i
\right)
=\sum_{n_i\in\Z} \delta(\lambda_i/g_s-n_i). 
\ee
The summation over integers $n_i$ is unconstrained, 
but the the case of $n_i=n_j$ drops from the above summation
due to the factor $\prod_{i<j}(n_i-n_j)^{2-2G}$ if $G=0$. 
For $G>1$, the partition function diverges at $n_i=n_j$, 
but we also simply drop these singular terms in the summation to
obtain regular results.
Therefore, we can assume that $n_i$'s are in a set of
strongly decreasing integer sequences $n_1 > n_2 > \cdots > n_N$
by using the Weyl permutation, too.
Thus we finally get a discretized version of the random matrix model
with the quadratic potential,%
\footnote{
In \cite{Gross:1993tu}, it is already suggested that 
the partition function of 
the ordinary two-dimensional Yang-Mills theory can be regarded as 
that of a discrete version of the random matrix model.} 
\be
Z=\sum_{n_1>n_2>\cdots>n_N}
\prod_{i < j}(g_s n_i-g_s n_j)^{2-2G}
\exp\left[
-\frac{g_s \mu A}{2}\sum_i n_i^2
\right].
\label{Gaussian discrete matrix model}
\ee
The difference from the ordinary random matrix model is that
the integral over eigenvalues is replaced by the
summation over possible sets of integer sequences.

So far, we have ignored the normalization factor of the above path
integral. In order to determine it,  
we require that a ``ground state'' configuration give $Z=1$.
The ground state configuration must be a densest sequence
of $n_i$ and satisfy $\sum_i n_i = 0$
(setting the ``origin'' of sequence at zero)
since the quadratic potential purposes to gather the eigenvalues $n_i$ at
the origin as much as possible. 
The densest sequence means each difference between neighbors is 1,
and from the condition $\sum_i n_i = 0$, we have $n_i=\frac{N+1}{2}-i$.
Plugging back this configuration into the sum, the ground state 
contributes to the partition function as 
\bea
Z_0 &=& \prod_{i<j}(g_s i-g_s j)^{2-2G}
e^{-\frac{g_s\mu A}{2}\sum_i(\frac{N+1}{2}-i)^2}\nn\\
&=& \prod_{i<j}(g_s i-g_s j)^{2-2G}
e^{-\frac{g_s\mu A}{2}\frac{N(N^2-1)}{12}}. 
\eea
This normalization factor itself has an important meaning since it is
proportional to the volume of $U(N)$ \cite{Ooguri:2002gx}.
Therefore, the partition function including the normalization factor 
becomes 
\be
Z=\sum_{n_1>n_2>\cdots>n_N}
\prod_{i < j}\left(\frac{n_i-n_j}{i-j}\right)^{2-2G}
\exp\left[
-\frac{g_s \mu A}{2}\left(
\sum_i n_i^2
-\frac{N(N^2-1)}{12}
\right)
\right].  
\ee

In order to make clear the relation to Migdal's partition function, 
where the partition function is expressed as a summation 
over irreducible representations of the gauge group $U(N)$, 
we rewrite the above partition function further.
The strongly decreasing sequence $n_i$ can be represented by a
weakly decreasing integer sequence
$k_1 \geq k_2 \geq \cdots \geq k_N$
through a relation, 
\be
n_i = k_i - i + c,  
\ee
with a constant $c$, 
since these $n_i$'s always satisfy
\be
n_i-n_j = k_i-k_j - i +j > 0, 
\ee
for any $i<j$. 
Here we can regard $k_i$ as the number of boxes in $i$-th row of 
a Young diagram corresponding to an irreducible representation 
of $U(N)$.

Using the parametrization $\{k_i\}$, 
the partition function reduces to
\be
Z=\sum_{\{k_i\}}
\prod_{i < j}\left(\frac{k_i-k_j-i+j}{i-j}\right)^{2-2G}
\exp\left[
-\frac{g_s \mu A}{2}\left(
\sum_i \left(Nk_i+k_i(k_i-2i+1)\right)
\right)
\right].
\ee
Notice that the dimension of the representation $R$ associated
with the Young diagram with $\{k_i\}$ is given by
\be
\dim R = \prod_{1 \leq i < j \leq N}\frac{k_i-k_j-i+j}{i-j},
\ee
and the quadratic Casimir is
\be
C_2(R) = \sum_i(Nk_i+k_i(k_i-2i+1)).
\ee
Thus we finally get Migdal's partition function, 
\be
Z=\sum_{R}
(\dim R)^{2-2G}
e^{
-\frac{\lambda A}{2N}
C_2(R)
},
\ee
where $\lambda=g_s \mu N$ is the 't Hooft coupling constant.
Thus we now understand that the discrete matrix model with the quadratic
potential and Migdal's partition function are related with each other by
a simple variable change, and both describe the non-perturbative 
dynamics of the two-dimensional Yang-Mills theory.

\begin{figure}[t]
\begin{center}
\includegraphics[scale=0.4]{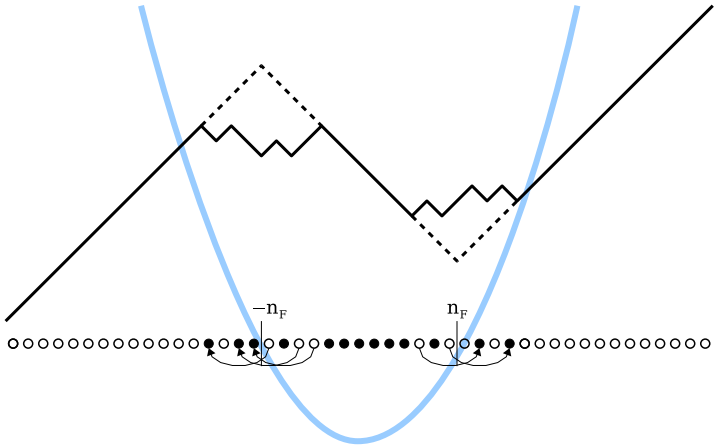}
\end{center}
\caption{An eigenvalue distribution in the quadratic potential.
Black dots stand for the positions of the eigenvalues and
white ones are vacant positions. If we draw the right-down and
right-up line for the black and white dots, respectively,
we obtain the Young diagram drawing above the dots.
The $n_F$ and $-n_F$ mean the positions of two fermi surfaces.}
\label{Maya diagram}
\end{figure} 
Let us finally discuss the behavior of the eigenvalues $\{n_i\}$ 
for the gaussian model. 
We first identify the positions of the eigenvalues $n_i$
with the so-called Maya diagram. 
The correspondence between the eigenvalues 
of the discrete matrix model and a Young diagram
is depicted in Fig.~\ref{Maya diagram}.
We also regard the positions of the eigenvalues 
as the Fock space of free fermions. 
The states of the free fermions have two fermi surface.
The fermions are exciting from the fermi surfaces
and the corresponding Young diagram is scraped away (melting) from the corner.
If we determine the constant $c$ by a condition
$\sum_i n_i = \sum_i k_i$, we have $c=\frac{N+1}{2}$ and the fermi
surfaces exist at $\frac{N-1}{2}$ and $-\frac{N-1}{2}$.
The ground state distribution is symmetric with respect to the critical
point of the quadratic potential, which is now at the origin.

\begin{figure}[t]
\begin{center}
\begin{tabular}{cc}
\includegraphics[scale=0.35]{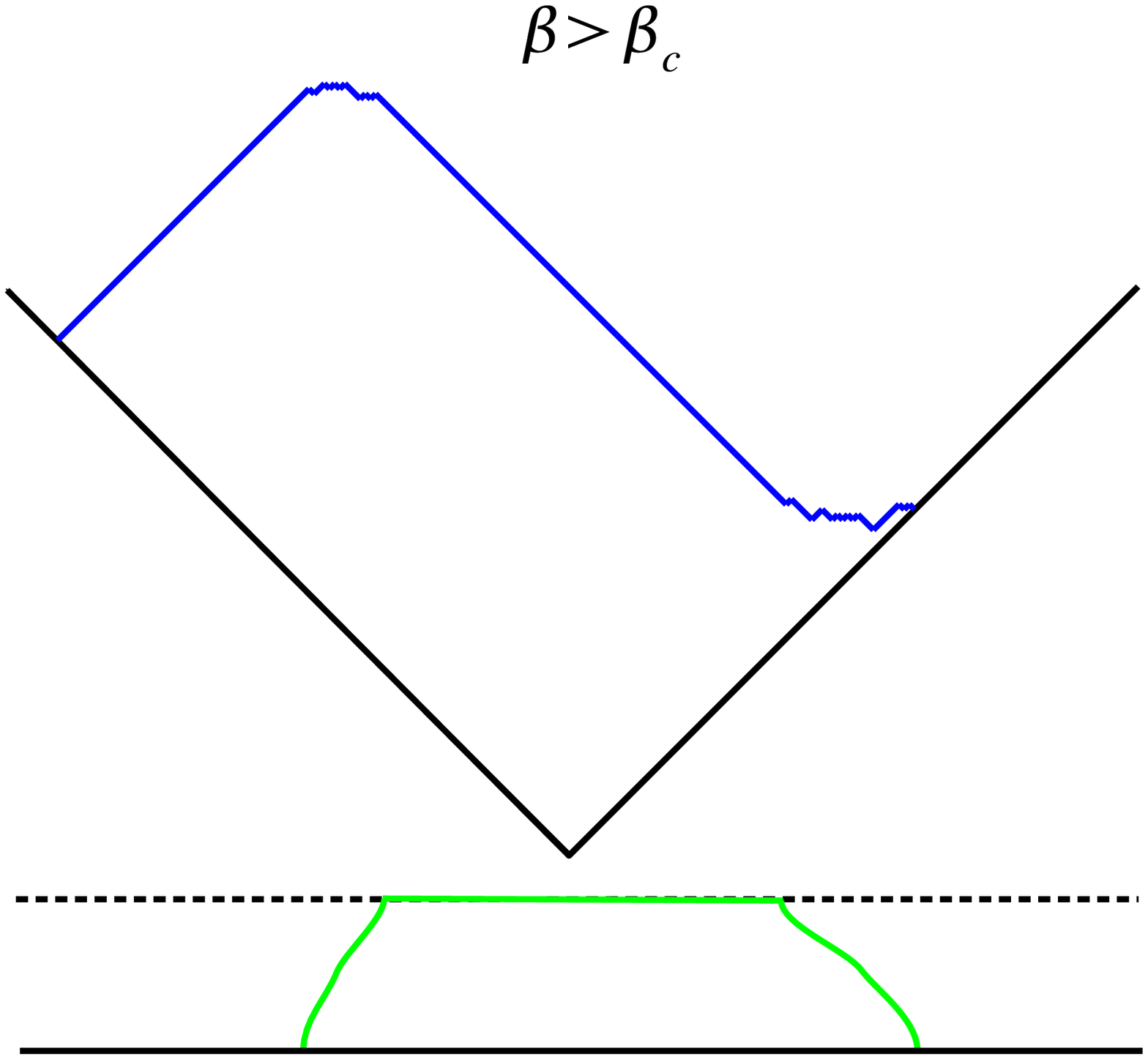}&
\includegraphics[scale=0.35]{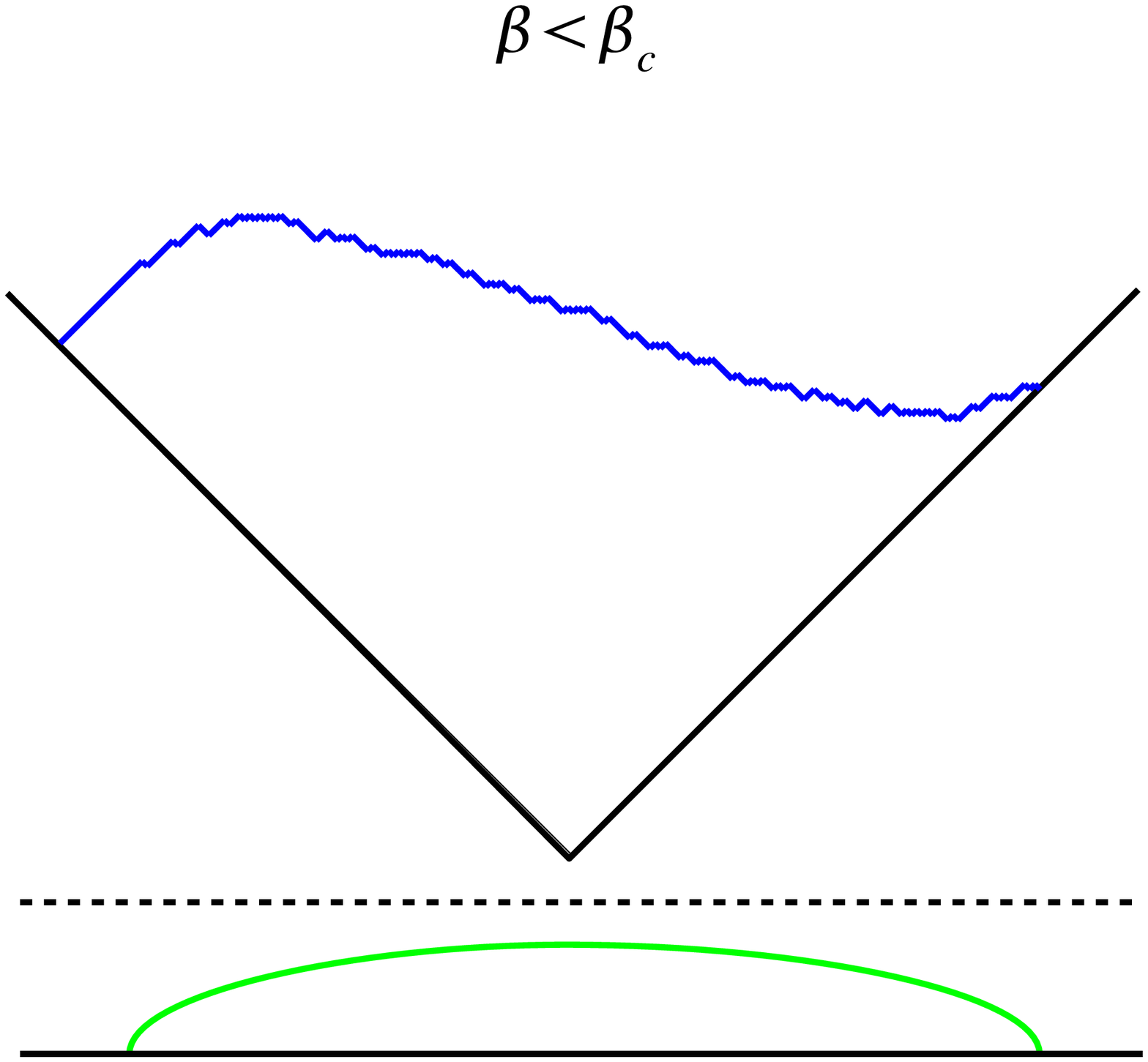}
\end{tabular}
\end{center}
\caption{These are results from
Monte-Carlo simulations for $N=100$ and $G=0$. 
When $\beta>\beta_c$, the
 eigenvalues are accumulated around the origin exceeding the maximal
 value of the density. The corresponding Young diagram has a planar
 edge. If the inverse temperature is below the critical value $\beta_c$,
 the system enters a different phase. The eigenvalue density
 behaves as the semi-circle.
}
\label{Monte-Carlo}
\end{figure} 
This choice or assumption for the ground state is also 
supported by the
numerical analysis of the discrete matrix model
(\ref{Gaussian discrete matrix model}) 
by using Monte-Carlo simulation.
For the symmetric quadratic potential,
the ground state is realized in the zero temperature limit
where the potential term is dominated
and the eigenvalues are symmetrically distributed.
We can also see the fermionic eigenvalues are excited from
two fermi surface if the repulsive force originated in the
Vandermonde determinant is taken into account. 
We draw some results of the Monte-Carlo simulation for $G=0$ in
Fig.\ref{Monte-Carlo}. We vary a combination of the parameters
$\beta\equiv g_s\mu A$,
which can be regarded as the inverse temperature of the system. 
In the low temperature (large $\beta$) limit, 
we see that the eigenvalues gather at the origin, namely
it approaches to the ground state configuration. 
The corresponding Young diagram is rigid 
and has a long planar edge of a rectangle shape. 
As the temperature becomes higher, 
the eigenvalues leave from the two fermi surfaces 
because of the repulsive effect from the Vandermonde determinant. 
Correspondingly, the Young diagram is crumbling (melting) from the corner.
On the other hand, 
if we define a density of the eigenvalue distribution, 
\begin{equation}
 \rho(x) = \frac{1}{N}\sum_{i=1}^{N}\delta(x-g_sn_i), 
\end{equation}
we find that there exists a maximal value of the density 
reflecting the discreteness of the eigenvalues. 
A region where the eigenvalue density meets the maximal value 
corresponds to a planar (right-down) edge of the Young diagram. 
Looking at the behavior of the eigenvalue distribution (density) 
or the shape of the Young diagram, 
we notice a phase transition at a critical temperature. 
In fact, if the temperature becomes higher, 
the planar region of the Young diagram finally disappears. 
From the eigenvalue density point of view, 
the flat head which hits the maximal value of the density 
is pinched at the critical temperature 
and the so-called Wigner's semi-circle 
is realized in the high temperature phase. 
This type of the phase transition is known as the
Douglas-Kazakov phase transition of third order
\cite{Douglas:1993ii}
by considering a continuum limit of 
Migdal's partition function or discrete matrix model. 
We will discuss it in detail in section 5.

\subsection{Nekrasov's partition function}
\label{sec:nekr-patit-funct}

Here we derive the Nekrasov's instanton partition function
from $\gYM$
using the similar technique in the previous subsection.

The partition function of the $\gYM$
is
\be
Z=\int {\cal D}A {\cal D}\Phi
\exp\left[
-\frac{1}{g_s}\int_{\Sigma_G}
\Tr \left(
i\Phi F
+W(\Phi)\omega\right)
\right].
\label{gYM2}
\ee
To make the eigenvalues of $\Phi$ localize at suitable
positions, we choose that critical points of the potential $W(\Phi)$
exist at $x=q_l$, namely, 
\be
W'(x) = \mu_{r+1}\prod_{l=1}^{r}(x-q_l). 
\label{superpotential}
\ee
The parameters $q_l$ will be related with the moduli parameters
in four-dimensional $SU(r)$ Yang-Mills theory by a rescaling
and we finally allow each $q_l$ to have a complex value 
by an analytic continuation.

The $\gYM$ (\ref{gYM2}) can be solved exactly
by using the previous procedure for Migdal's partition function.
The result is
\be
Z=\sum_{\{n_i\}}
\prod_{1 \leq i < j \leq N}\left(g_sn_i-g_sn_j\right)^{2-2G}
e^{
-\frac{A}{g_s}\sum_i W(g_s n_i)
},
\label{discrete MM}
\ee
up to a normalization factor. So the partition function of 
$\gYM$ can be regarded as a discretized version 
of the random matrix model with a generic potential (action).
In the case of $\mu_{r+1}\gg 1$, 
the above summation over $n_i$ 
dominates near the critical points $q_l$,
so we can assume that $N_l$ eigenvalues are around 
the critical point $q_l$ as 
\begin{multline}
g_s n_i = (\underbrace{q_1+g_s h^{(1)}_1,\ldots,q_1 + g_s h^{(1)}_{N_1}}
_{N_1},
\underbrace{q_2+g_s h^{(2)}_1,\ldots,q_2+ g_s h^{(2)}_{N_2}}_{N_2},\\
\ldots,
\underbrace{q_r+g_s h^{(r)}_1,\ldots,q_r+ g_s h^{(r)}_{N_r}}_{N_r}),
\label{fluctuation}
\end{multline}
where $h^{(l)}_i$ $(i=1,\ldots,N_l)$ 
represent fluctuations around the critical points, 
which satisfy 
\be
h^{(l)}_1 > h^{(l)}_2 > \cdots > h^{(l)}_{N_l},
\ee
for each $l$.
The $U(N)$ gauge symmetry is broken to $\prod_{l=1}^{r} U(N_l)$
around this configuration. 
In contrast to the previous Gaussian case (quadratic potential),
we can not say the eigenvalues are distributed symmetrically around each
critical point in general
since we can not ignore effects from neighbor critical points.
However, in the large $N$ limit which we will discuss later, 
the generic
potential decouples into a union of the quadratic potentials in an
approximation. So we hereafter proceed our analysis by assuming
the eigenvalues are symmetrical at each critical point,  
although this is valid only for a suitable scaling large $N$ limit.

Under this assumption,
we introduce a parametrization $\{K_i^{(l)}\}$ as  
\be
h^{(l)}_i = K^{(l)}_i - i +\frac{N_l+1}{2},
\ee
and we only consider the case 
$\sum_{i=1}^{N_l}g_s K_i^{(l)}\ll |q_l-q_n|$ 
for ${}^\forall n \ne l$ 
so that the lumps of the eigenvalues separate enough each other. 
Here we have assumed $N_l$'s are odd numbers for simplicity.  
As well as the argument in the previous subsection, 
$\{K_i^{(l)} \}$ is weakly decreasing integer sequence; 
\be
K^{(l)}_1 \geq K^{(l)}_2 \geq \cdots \geq K^{(l)}_{N_l},  
\ee
and we call the densest configuration $\{K_i^{(l)}=0\}$ 
as the ``ground state'' of this system.

\begin{figure}[t]
\begin{center}
\begin{tabular}{rcl}
\includegraphics[scale=0.5]{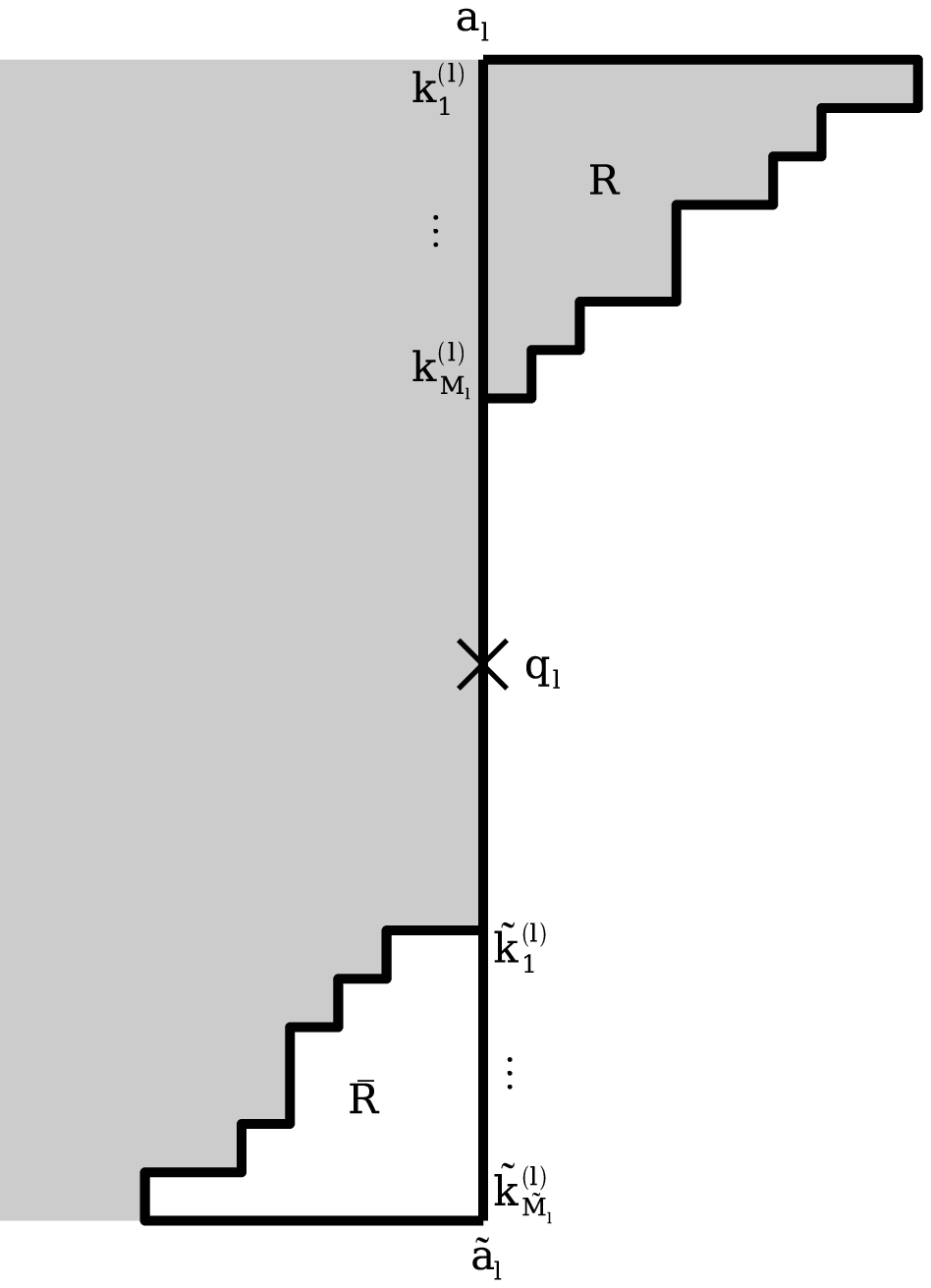}&
\hspace*{0.5cm}&
\includegraphics[scale=0.5]{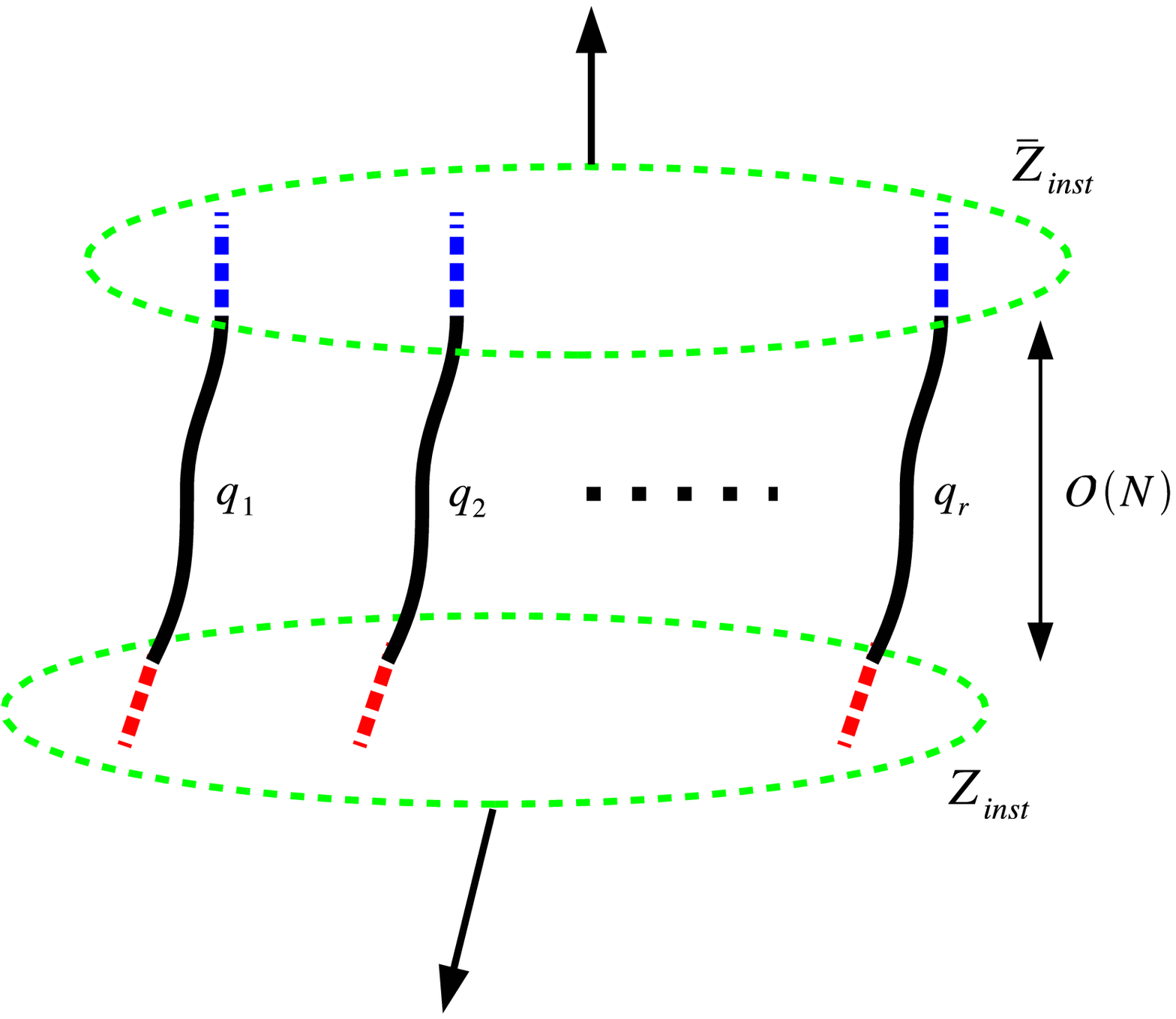}
\end{tabular}
\end{center}
\caption{We decompose the excitations from two fermi surfaces. We denote
 these excitations as $k_i^{(l)}$ and $\tilde{k}_i^{(l)}$ and positions
 of the surfaces as $a_l$ and $\tilde{a}_l$. In the
 large $N$ limit, the two surfaces are decoupled with each other. And
 one surface contributes to the instanton partition function.
}
\label{chiral decomp}
\end{figure}

Let us next consider excitations from the ground state, 
that is, non-trivial configurations of $\{K_i^{(l)}\}$. 
Since we assume symmetrical distribution of eigenvalues 
around each critical point, 
$\{K_i^{(l)}\}$ is assumed to be separated into 
two parts (positive $K^{(l)}_i$ and negative $K^{(l)}_i$) as%
\footnote{As seen in \cite{Matsuo:2004cq}, this type of the
limiting shape of the Young diagram is dominated in the
Douglas-Kazakov phase ($\beta>\beta_c=\pi^2$).}
\begin{equation}
K_i^{(l)} = 
\begin{cases}
k_i^{(l)} \ge 0, & (i=1,\cdots,\frac{N_l-1}{2}) \\
-\tk_{N_l-i+1}^{(l)} \le 0  & (i=\frac{N_l+3}{2},\cdots,N_l), 
\end{cases}
\end{equation} 
where both of $\{k_i^{(l)}\}$ and $\{\tk_j^{(l)}\}$ are 
weakly decreasing sequences of non-negative integer. 
(We assume $K_{(N_l+1)/2}^{(l)}=0$.) 
If we  consider the case that 
the numbers of the excitations are much smaller than $N_l$, 
most of $\{k_i^{(l)}\}$ and $\{\tk_j^{(l)}\}$ are zero 
and the non-zero elements express excitations from ``fermi surfaces'' 
at $a_l \equiv q_l+g_s\frac{N_l-1}{2}$ and
$\tilde{a}_l \equiv q_l-g_s\frac{N_l-1}{2}$.
(See Fig.\ref{chiral decomp}.)
Note that 
this decomposition corresponds to the chiral decomposition
discussed in \cite{Gross:1993hu, Gross:1993yt, Gross:1993cw}. 

Using these notations, the potential part of the partition
function becomes
\begin{align}
\frac{A}{g_s}\sum_{i=1}^{N}W(g_s n_i)
&=\frac{A}{g_s}\sum_{l=1}^{r}\sum_{i=1}^{\frac{N_l-1}{2}}
\left[
W\left(a_l +g_s(k^{(l)}_i-i+1)\right)
+W\left(a^-_l-g_s(\tilde{k}^{(l)}_i-i+1)\right)
\right] \nn \\
& \simeq 
\frac{A}{g_s}\sum_{l=1}^{r}\sum_{i=1}^{\frac{N_l-1}{2}} 
\Bigl[
W\left(a_l + g_s(-i+1)\right) 
+ W'\left((a_l + g_s(-i+1)\right)g_sk_i^{(l)} \nn \\
&\hspace{28mm}+W\left(\widetilde{a}_l - g_s(-i+1)\right) 
- W'\left((\widetilde{a}_l - g_s(-i+1)\right)g_s\tk_i^{(l)} 
\Bigr], 
\label{potential-pre}
\end{align}
where we have expanded each term by ``small'' fluctuations 
$g_sk^{(l)}_i$ and $g_s \tilde{k}^{(l)}_i$. 
Note that $k_i^{(l)}$ and $\tk_i^{l}$ have non-zero values 
only for $i\ll N_l$, where the arguments of $W'$ in 
(\ref{potential-pre}) can be evaluated as 
\begin{equation}
a_l + g_s(-i+1) \simeq \frac{g_s N_l}{2}+\cO(1). 
\end{equation}
Therefore, from the definition of the potential (\ref{superpotential}), 
we see that the potential part of the partition function 
is evaluated as 
\begin{align}
\exp\left[-\frac{A}{g_s}\sum_{l=1}^r
\sum_{i=1}^{N_l}W(g_sn_i^{(l)})\right] 
&\simeq C_0\exp\left[
-\mu_{r+1}A\sum_{l=1}^{r}\left(\frac{g_s N_l}{2}\right)^r
(k_l+\tk_l)
\right], 
\label{potential part}
\end{align}
where $C_0$ is the contribution from the ground state; 
$C_0=e^{-\frac{A}{g_s}\sum_{l=1}^r\sum_{i=1}^{N_l}W
\left(a_l+g_s(-i+\frac{N_l+1}{2})\right)}$, 
and we have used a suitable analytic continuation 
in the anti-chiral sector.

On the other hand, the discrete Vandermonde determinant
(measure) part of the partition function is decomposed as 
\be
\prod_{i\neq j} (g_s n_i - g_s n_j)=
\mu_{++}^{(N)}
\mu_{+-}^{(N)}
\mu_{-+}^{(N)}
\mu_{--}^{(N)}
\ee
where
\bea
\mu_{++}^{(N)} &=&
\prod_{(l,i_l)\neq(n,j_n)}^{\{N_{l,n}/2\}}
\left(
{a_l-a_n+g_s(k^{(l)}_{i_l}-k^{(n)}_{j_n}-i_l+j_n)}
\right),\\
\mu_{+-}^{(N)} &=&
\prod_{(l,i_l)\neq(n,j_n)}^{\{N_{l,n}/2\}}
\left(
{a_l-\widetilde{a}_n+g_s(k^{(l)}_{i_l}+\tilde{k}^{(n)}_{j_n}-i_l-j_n+1)}
\right),\\
\mu_{-+}^{(N)} &=&
\prod_{(l,i_l)\neq(n,j_n)}^{\{N_{l,n}/2\}}
\left(
{\widetilde{a}_l-a_n-g_s(\tilde{k}^{(l)}_{i_l}+k^{(n)}_{j_n}-i_l-j_n+1)}
\right),\\
\mu_{--}^{(N)} &=&
\prod_{(l,i_l)\neq(n,j_n)}^{\{N_{l,n}/2\}}
\left(
{\widetilde{a}_l-\widetilde{a}_n-g_s(\tilde{k}^{(l)}_{i_l}-\tilde{k}^{(n)}_{j_n}-i_l+j_n)}
\right), 
\eea
where the symbol $\{N_{l,n}/2\}$ over the product represents 
that each $i_l$ ($j_n$) runs from 1 to $\frac{N_l-1}{2}$ ($\frac{N_n-1}{2}$),  
and we have used the suffix $(N)$ to stress that 
they are defined for finite $N$. 
Here we find that 
the cross terms $\mu_{\pm\mp}^2$ of are decoupled
from the partition function in the large $N$ limit 
since $a_l-a_n \sim {\cal O}(1)$, $\ta_l-\ta_n\sim\cO(1)$
and $a_l-\ta_n  \sim {\cal O}(g_s N)$. 



Now we would like to take a large $N$ limit. 
In our case, we must also take the limit of $\mu_{r+1}\to 0$ 
to recover $\cN{=}2$ SUSY in the four-dimensional space-time, 
thus it is proper to take the large $N$ limit with 
$N_l=N_n$ for all $l=1,\cdots,r$ 
since the positions of the critical points are in the flat direction 
of the $\cN{=}2$ SUSY theory.   
Then, we set $N_l$ as 
\begin{equation}
 N_l\equiv 2M+1, \qquad (l=1,\cdots,r)
\label{large N_l}
\end{equation}
and take the limit $M\to\infty$. 
In this limit, $\mu_{++}^{(N)}$ can be evaluated as 
\begin{align}
\mu_{++}^{(N)} &= Z_0(\av)
\prod_{(l,i_l)\neq(n,j_n)}^M
\left(
\frac
{a_l-a_n +g_s(k^{(l)}_{i_l}-k^{(n)}_{j_n}-i_l+j_n)}
{a_l-a_n +g_s(-i_l+j_n)}
\right) \nn \\ 
&\simeq Z_0(\av) (g_sM)^{2rk} 
\prod_{(l,i)\neq(n,j)} 
\left(
\frac
{a_l-a_n +g_s(k^{(l)}_{i}-k^{(n)}_{j}-i+j)}
{a_l-a_n +g_s(-i+j)}
\right), 
\label{Z+}
\end{align}
with 
\begin{align}
Z_0(\av) &= \prod_{(l,i)\ne(n,j)}\Bigl(
a_l-a_n +g_s(-i+j) \Bigr), 
\label{perturbative}
\end{align}
where $k=\sum_{l=1}^r\sum_{i}k_{i}^{(l)}$ and 
$i$ and $j$ now run from $1$ to $\infty$. 
For the proof from the first line to the second line 
of (\ref{perturbative}), see the appendix A.
This normalization factor is coming from a contribution of the ground
state to the Vandermonde measure, and will give the perturbative part of
the prepotential of the four-dimensional $\N{=}2$ SUSY gauge theory
\cite{Losev:2003py,Nekrasov:2003rj,Maeda:2004iq}.
In addition, we take a scaling limit $\mu_{r+1}\to 0$ 
with fixing,  
\begin{equation}
(g_sM)^{2r} \exp\left(-A \mu_{r+1}(g_s M)^r\right) \equiv \Lambda^{2r}. 
\label{double scaling limit} 
\end{equation}
On the other hand, 
we can also show that $\mu_{--}$ gives the same contribution with
$\mu_{++}$ under the setup (\ref{large N_l}) 
since $\ta_l-\ta_n$ and $a_l-a_n$ coincide 
in the large $N$ limit (\ref{large N_l}). 
Therefore, in this double scaling limit,  
the partition function of the $\gYM$ becomes 
\be
{Z}_{gYM_2} = 
Z_{\rm inst}\left(\av;\Lambda \right)^2
\label{gYM2 to Nekrasov}
\ee
with 
\begin{align}
Z_{\rm inst}\left(\av;\Lambda \right) &= 
Z_0(\av)\sum_{\vec{\kv}}
\Lambda^{2rk}
\prod_{(l,i)\neq (n,j)}
\left(
\frac{a_l-a_n+g_s(
k^{(l)}_i-k^{(n)}_j+j-i)}
{a_l-a_n+g_s(
j-i)}
\right)^{1-G}. 
\end{align}
In particular, 
if we set $G=0$, $Z_{\rm inst}$ coincides with the Nekrasov's partition
function exactly. 

Finally we note that, 
if we introduce the $\theta$-term in the action of the gYM$_2$, 
we can easily show that the ``QCD scale'' $\Lambda$ 
in (\ref{double scaling limit}) becomes complex 
and $\bar{\Lambda}$ appears in $\mu_{--}$. 
Then, we expect that $\mu_{++}$ and $\mu_{--}$ describe 
the instanton and the anti-instanton contributions to the prepotential,
respectively, and we could verify 
\begin{equation}
 Z_{gYM_2} = \left|Z_{\rm inst}(\av;\Lambda)\right|^2,  
\end{equation}
by taking a suitable analytic continuation of $\av$.

\section{Adding matters}
\label{sec:adding-matters}

In this section, we derive the instanton partition function  
for other $\cN{=}2$ theories from suitable two-dimensional theories. 
We first consider the $\cN{=}2$ theory with 
massive hypermultiplets in the adjoint representation, 
and next we present a formula for quiver gauge theory 
\cite{Fucito:2004gi} as a conjecture.  
We also show that a formula for theory  
with fundamental hypermultiplets given in \cite{Nekrasov:2003rj} 
is obtained by a flow from the quiver theory. 

\subsection{Addition of adjoint matter}
\label{sec:addit-adjo-matt}

To construct a two-dimensional theory that provides 
the instanton counting of the $\cN{=}2$ theory with the massive 
adjoint hypermultiplet, 
we first introduce an additional BRST multiplets 
$(X,\psi_X)$ and $(Y,\psi_Y)$ whose (deformed) 
BRST transformation is given by 
\be
\begin{array}{ll}
Q X = \psi_X, & Q \psi_X = [\Phi,X]+m X, \\
Q Y = \psi_Y, & Q \psi_Y = [\Phi,Y]-m Y,
\end{array}
\ee
where $X$ and $Y$ are both Hermitian
in the adjoint representation of $U(N)$. 
Using them, we add the following BRST exact expression to the 
action of the topological field theory (\ref{topological action}), 
\begin{align}
\delta S_{\rm adj} 
&= \frac{1}{2g_s} \int_{\Sigma_G} 
Q \left(Y \psi_X-X\psi_Y\right) \nn \\
&= \frac{1}{g_s} \int_{\Sigma_G} \left(
\Phi[X,Y] + m X Y - \psi_X \psi_Y\right), 
\end{align} 
and consider the expectation value in this topological field theory, 
\begin{equation}
\left\langle
\exp\left[ -\frac{1}{g_s}\int_{\Sigma_G} 
\Tr\left(
i\Phi F + \frac{1}{2}\lambda\wedge\lambda + 
W(\Phi)\omega
\right) 
\right]
\right\rangle_{\rm top}, 
\label{adjoint top}
\end{equation}
where $W(\Phi)$ is again defined by (\ref{superpotential}). 
As discussed in the section 2, it can be 
evaluated as the partition function of a corresponding 
two-dimensional gauge theory, 
\be
Z_{\rm adj} = \int \D A \D \Phi \D X \D Y
\exp\left\{
-\frac{1}{g_s}
\int_{\Sigma_G}\, \Tr\Bigl[
i\Phi F  + \bigl(
W(\Phi) + \Phi[X,Y] + m XY\bigr)\omega
\Bigr]
\right\}, 
\label{adjoint partition}
\ee
after integrating out $\lambda$, $\psi_X$ and $\psi_Y$,
up to contributions from higher critical points.
We assume that these ``matter'' fields are localized at several points
 as impurities
(sources) on the Riemann surface $\Sigma_G$ as well as the Hitchin system in
 \cite{Kapustin:1998pb}.

In the following, we show that (\ref{adjoint partition}) 
reproduces the instanton counting of the four-dimensional 
$\cN{=}2$ theory with the massive hypermultiplet 
in the adjoint representation given in \cite{Bruzzo:2002xf,Nekrasov:2003rj}. 
Fixing the gauge by $\Phi_\alpha=0$, 
as discussed in the section \ref{sec:migd-part-funct}, 
we obtain (\ref{discrete MM}) by performing the Gaussian integrals  
over the gauge fields $A_\alpha$ and the ghost fields 
$(c_\alpha,\bar{c}_\alpha)$. 
In the case of (\ref{adjoint partition}), 
we must also perform the integration over $X$ and $Y$, 
\begin{equation}
\int \prod_{i\ne j}^N \cD X_{ij} \D Y_{ij}
\exp\left\{
-\frac{A}{g_s}
\sum_{i\ne j}^N \left(m+\lambda_i-\lambda_j\right)X_{ij}Y_{ij}
\right\}
=\prod_{i\ne j}^N \frac{1}{m+\lambda_i-\lambda_j},
\end{equation}
where we use the assumption that $X$ and $Y$ are delta-functionally
localized at a point on $\Sigma_G$.
Then, except for an overall factor, 
(\ref{adjoint partition}) can be evaluated as 
\begin{equation}
Z_{\rm adj}=\sum_{n_i \in \Z} 
\prod_{i\neq j}^N
\frac{\left(g_s n_i- g_s n_j\right)^{1-G}}{m+g_s n_i- g_s n_j}
e^{-\frac{A}{g_s}\sum_{i=1}^N W(g_s n_i)}. 
\end{equation}

Now we take the double scaling limit $N\to\infty$ and $\mu_{r+1}\to 0$
with the condition (\ref{double scaling limit}). 
Repeating the discussion in the section \ref{sec:nekr-patit-funct},
we can show that 
the partition function (\ref{adjoint partition}) becomes
\begin{equation} 
 Z_{\rm adj} = 
Z_{\rm inst}^{\rm adj}
\left(\av,m;\Lambda \right)^2
\end{equation}
with 
\begin{align} 
Z_{\rm inst}^{\rm adj} &= 
Z_{\rm pert}^{\rm adj}(\av,m)
\sum_{\vec{\kv}}\Lambda^{2rk}
\prod_{(l,i)\neq (n,j)}
\left(\frac{a_l-a_n+g_s(k^{(l)}_i-k^{(n)}_j+j-i)}
{a_l-a_n+g_s(j-i)}
\right)^{1-G} \nn \\
&\hspace{3.5cm}\times\frac{m+a_l-a_n+g_s(j-i)}
{m+a_l-a_n+g_s(k^{(l)}_i-k^{(n)}_j+j-i)},\label{chiral-adjoint}
\end{align}
where 
\begin{align}
Z_{\rm pert}^{\rm adj}(\av,m) &= 
\prod_{(l,i)\neq (n,j)} 
\frac{\left(a_l-a_n+g_s(j-i)\right)^{1-G}}
{m+a_l-a_n+g_s(j-i)}. 
\label{perturbative-adjoint}
\end{align} 
If we set $G=0$, the chiral part $Z_{\rm inst}^{\rm adj}$ reproduces 
the instanton counting for 
the $\cN{=}2$ theory with the massive adjoint hypermultiplet. 
Here we have again used the fact that $\ta_l-\ta_n$ can be 
replaced with $a_l-a_n$ under the condition (\ref{large N_l}).

\begin{figure}[t]
\begin{center}
\begin{tabular}{ccc}
\includegraphics[scale=0.5]{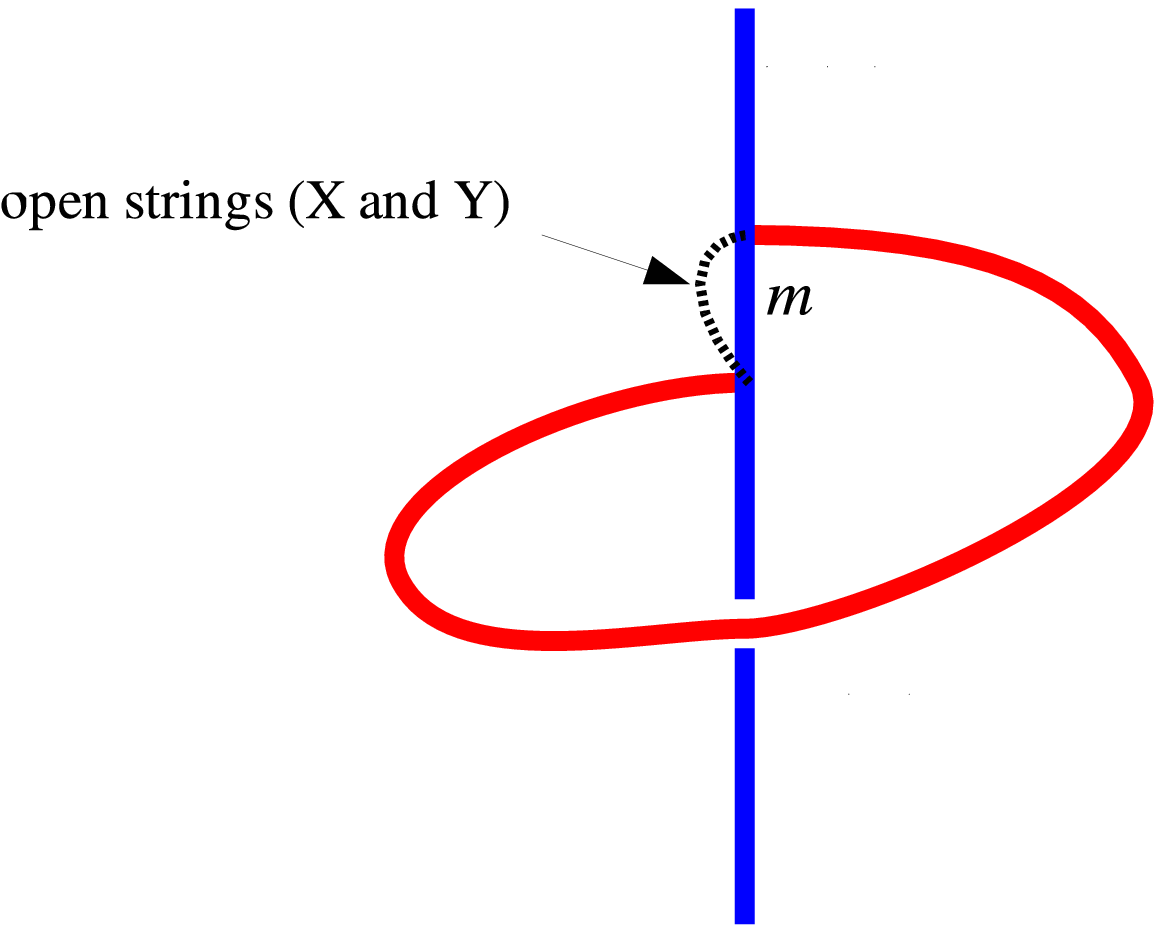}&
\hspace*{0.5cm}
$\begin{array}{c}
\Longrightarrow\\
m\rightarrow 0
\end{array}$
\hspace{1.5cm}&
\includegraphics[scale=0.5]{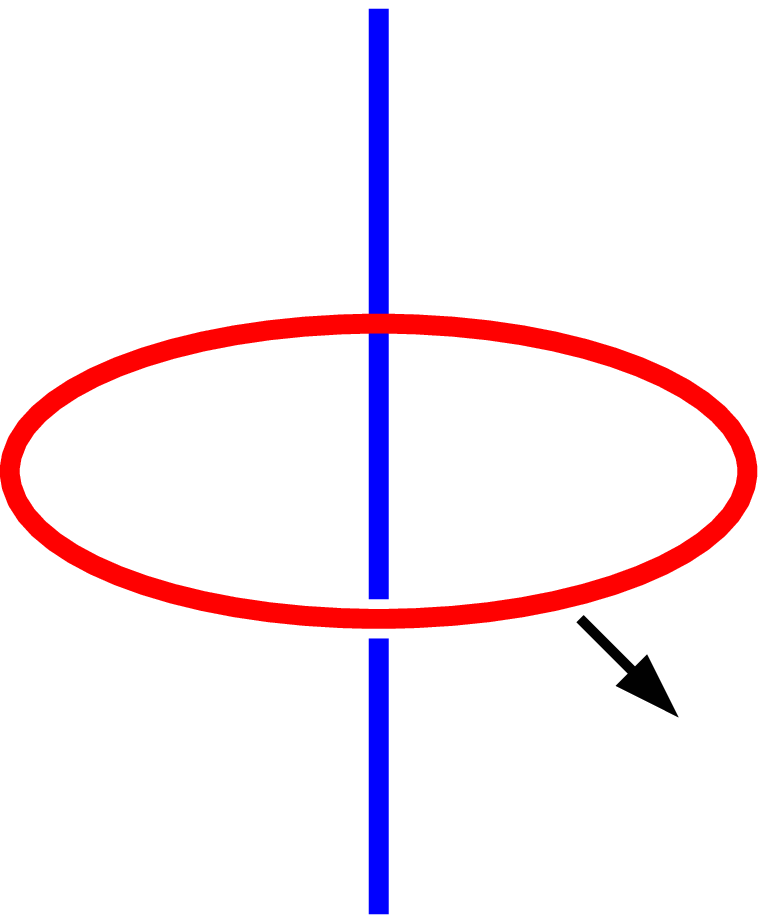}\\
(a) & & (b)
\end{tabular}
\end{center}
\caption{A brane configuration of $\N{=}2$ theory with adjoint
 hypermultiplets. D4-branes are wrapping around a circle with a
 shift of the end points on a NS5-brane. The shift disappears in the
 massless limit and the D4-branes can be removed from the NS5-brane.
T-duals of these brane configurations explain why the effective genus in
 the instanton partition function is increased by one.}
\label{massless limit}
\end{figure} 

We would like to comment on the massless limit of the above model. 
If we see the expressions (\ref{chiral-adjoint}) and
(\ref{perturbative-adjoint}), we notice that 
the contribution from $X$ and $Y$ cancels one of the $(1-G)$ 
contributions from $\gYM$ in the limit of $m\to 0$. 
This means that the genus $G$ of the Riemann surface effectively shifts
to $G+1$. This phenomenon can be easily understood by using a brane
configuration for the $G=0$ case.
The brane configuration for $\N{=}2^*$ (Donagi-Witten)
\cite{Donagi:1995cf} is shown in Fig.\ref{massless limit} 
where $N$ D4-branes are ending on a single
NS5-brane and wrapping around a compact circle. 
The end points of the D4-branes are separated by $m$. 
Open strings between two end points have
a minimal length and get mass proportional to $m$. They correspond to
the hypermultiplets $X$ and $Y$. In the massless limit, the theory enhances to
$\cN{=}4$ since the two adjoint hypermultiplets are put into the $\cN{=}4$
vector multiplet. This means the D4-branes can be removed from the
NS5-brane in the brane configuration and they are T-dual to parallel
D3-branes. 
T-dual of the brane configuration maps the D4-branes to the D5-branes
wrapping on $S^2$ and the NS5-brane to a deficit in the geometry.
On the other hand, the D4-branes wrapping the compact circle without any
shift is the T-dual of D5-branes wrapping on $T^2$ where the SUSY enhances to
$\cN{=}4$. Therefore, we expect that two punctures on $S^2$ corresponding
to the end points on the NS5-brane join with each
other and the topology changes from $S^2$ to $T^2$ in the massless
limit, namely, the genus of the cycle is increased by one.
This phenomenon should be extended to any genus $G$. If two punctures on
the Riemann surface with genus $G$ join in the massless limit, the genus
is increased by one. This picture agrees with our derived formula.

\subsection{Quiver theory}
\label{sec:quiver-theory}

In this succeeding subsection, we construct a two-dimensional theory 
which is thought to provide the instanton counting of the $\cN{=}2$
$A_2$-type quiver gauge theory. 
The generalization to an arbitrary $A_n$-type 
quiver theory should be straightforward. 

The $A_2$ quiver theory is realized on two Riemann surfaces which intersect
at a point. Bi-fundamental matters in the quiver gauge theory come from
open strings connecting D-branes wrapping on the Riemann surfaces
independently.
So let us consider the system with $N_1$ D-strings on Riemann surface 
$\Sigma_{G_1}$ with genus $G_1$ and 
$N_2$ D-strings on $\Sigma_{G_2}$ with genus $G_2$.  
As discussed in the section \ref{sec:localization-d-brane},  
the low energy effective theory of this system is a topologically
twisted theory constructed from BRST multiplets 
$(A_1,\lambda_1,\Phi_1)$ and $(A_2,\lambda_2,\Phi_2)$, 
$(\bar{\Phi}_1,\eta_1)$ and $(\bar{\Phi}_2,\eta_2)$, 
and $(H_1, \chi_1)$ and $(H_2, \chi_2)$, 
which are in the adjoint representation of $U(N_1)$ and $U(N_2)$, 
respectively. 
The BRST transformations are given as well as
(\ref{BRST1}) and (\ref{BRST2}), 
and the action is given by 
\begin{align}
S_{\rm top}=
Q\Bigl(
&\frac{1}{h_1^2}\int_{\Sigma_{G_1}} 
\Tr_{N_1}\Xi(A_1,\lambda_1,\Phi_1,\bar{\Phi}_1,\eta_1,H_1,\chi_1) \nn \\
&+\frac{1}{h_2^2}\int_{\Sigma_{G_2}} 
\Tr_{N_2}\Xi(A_2,\lambda_2,\Phi_2,\bar{\Phi}_2,\eta_2,H_2,\chi_2)
\Bigr). 
\label{pre quiver action}
\end{align}
where $\Tr_{N_a}$ denotes the trace over an $N_a\times N_a$ matrix, 
and the function $\Xi$ is defined like (\ref{Xi}). 
This theory is again independent of the coupling constants 
$h_1$ and $h_2$. 

As in section \ref{sec:localization-d-brane}, 
we first consider following observables of 
the topological field theory, 
\begin{equation} 
e^{-\frac{1}{g_s}I_2^{(a)}} \equiv 
\exp\left\{
-\frac{1}{g_s}\int_{\Sigma_{G_a}} \Tr_{N_a}\left(
i\Phi_aF_a + \frac{1}{2}\lambda_a\wedge\lambda_a
+W_a(\Phi_a)\omega_a\right)
\right\}, \qquad (a=1,2)
\label{instanton chemical potential}
\end{equation} 
where $\omega_a$ is the volume form on $\Sigma_{G_a}$ 
and we assume that the potentials $W_1(x)$ and $W_2(x)$ satisfy 
\begin{align}
W'_1(x) &= \mu_{r+1}(x-q^{(1)}_1)\cdots(x-q^{(1)}_r), \\
W'_2(x) &= \nu_{s+1}(x-q^{(2)}_1)\cdots(x-q^{(2)}_s), 
\end{align}
respectively. 
If we evaluate the expectation value of $e^{-\frac{1}{g_s}I_2}$ 
in the topological field theory,
we obtain the product of Nekrasov's partition functions 
corresponding to pure $\cN{=}2$ SYM theories with the gauge groups 
$SU(r) \times SU(s)$ after taking a suitable large $N$ limit.

Furthermore, we would like to construct another 
``potential term'' that
produces the quiver-type interaction in the four-dimensional
$\cN{=}2$ theory. 
To achieve this, we again introduce an additional BRST multiplets 
$(B,\psi_B)$ and $(\tB,\psi_\tB)$ which are $N_1\times N_2$ and 
$N_2\times N_1$ matrices, respectively. 
The BRST transformations are given by 
\begin{align}
QB &= \psi_B, \quad Q\psi_B = \Phi_1 B - B \Phi_2,   \\ 
Q\tB &= -\psi_\tB,\quad Q\psi_\tB = \tB \Phi_1 - \Phi_2 \tB, 
\label{adjoint BRST}
\end{align}
or, in component, 
\begin{align}
(QB)_{i_1i_2} &= (\psi_B)_{i_1i_2}, \quad 
(Q{\psi_B})_{i_1i_2} = \sum_{j_1=1}^{N_1}(\Phi_1)_{i_1j_1}B_{j_1i_2}
-\sum_{j_2=1}^{N_2}B_{i_1j_2}(\Phi_2)_{j_2i_2},   \\
(Q\tB)_{i_2i_1} &= -(\psi_\tB)_{i_2i_1}, \quad 
(Q{\psi_\tB})_{i_2i_1} = \sum_{j_1=1}^{N_1}\tB_{i_2j_1}(\Phi_1)_{j_1i_1}
-\sum_{j_2=1}^{N_2}(\Phi_2)_{i_2j_2}\tB_{j_2i_1}, 
\end{align}
where $i_1,j_1=1,\cdots,N_1$ and $i_2,j_2=1,\cdots,N_2$. 
We assume that these ``matter'' fields are also localized at points as
impurities.
Now let us define the topological field theory by adding the following
BRST exact term, 
\begin{align}
\delta S_{\rm quiv} &\equiv 
\frac{1}{2g_s} \int_{\Sigma_{G_1}\cup\Sigma_{G_2}} 
Q \Tr \left(\tB\psi_B
+B\psi_\tB \right) \nn \\
&= \frac{1}{g_s} \int_{\Sigma_{G_1}\cup\Sigma_{G_2}}
\left[
\Tr_{N_2}\left(\tB \Phi_1 B\right) - \Tr_{N_1}
\left(B\Phi_2 \tB \right)-\Tr_{N_2}\left(\psi_\tB\psi_B\right)
\right],
\end{align}
to the action (\ref{pre quiver action}).   
The instanton partition function of the quiver gauge theory can be
obtained from the evaluation of the expectation value, 
\be 
\left\langle
e^{-\frac{1}{g_s}\left(
I_2^{(1)}+I_2^{(2)} 
\right)}
\right\rangle_{\rm top}. 
\ee
If we ignore contributions from higher critical points again, 
it reduces to the partition function of a two-dimensional theory, 
\begin{align}
Z_{\rm quiv} = \int& \cD A_1 \cD\Phi_1 \cD A_2 \cD \Phi_2 \cD B \D \tB \nn \\
&\exp\Biggl\{
-\frac{1}{g_s}\int_{\Sigma_{G_1}} 
\left[\Tr_{N_1}
\left(i\Phi_1F_1 +W_1(\Phi_1)\omega_1 \right)
+\Tr_{N_2}\left(\tB \Phi_1 B\right)\omega_1\right] \nn \\
&\hspace{1cm}-\frac{1}{g_s}\int_{\Sigma_{G_2}} 
\left[\Tr_{N_2}\left(i\Phi_2F_2 +W_2(\Phi_2)\omega_2 \right)
-\Tr_{N_1}\left(B \Phi_2 \tB \right)\omega_2\right]
\Biggr\}, 
\label{quiver partition}
\end{align}
where we have used the assumption that 
$\Phi_1$ $(\Phi_2)$ takes non-zero value only on $\Sigma_1$ $(\Sigma_2)$. 
Repeating the discussion in the section \ref{sec:nekr-patit-funct}, 
the partition function (\ref{quiver partition}) can be expressed 
as a summation over sets of integers $\{n^{(1)}_{i_1}\}$ 
and $\{n^{(2)}_{i_2}\}$; 
\begin{align}
Z_{\rm quiv} = \sum_{n^{(1)}_{i_1},n^{(2)}_{i_2}\in \Z}
&\prod_{i_1\ne j_1}^{N_1}
\left(g_sn_{i_1}^{(1)}-g_sn_{j_1}^{(1)}\right)^{1-G_1}
\prod_{i_2\ne j_2}^{N_2}
\left(g_sn_{i_2}^{(2)}-g_sn_{j_2}^{(2)}\right)^{1-G_2} \nn \\
&\times\prod_{i_1=1}^{N_1}\prod_{i_2=1}^{N_2}
\left(g_sn_{i_1}^{(1)}-g_sn_{i_2}^{(2)}\right)^{-1}
e^{-\frac{A_1}{g_s}\sum_{i_1=1}^{N_1}W_1(g_sn_{i_1}^{(1)})}
e^{-\frac{A_2}{g_s}\sum_{i_2=1}^{N_2}W_2(g_sn_{i_2}^{(2)})}, 
\label{quiver partition2}
\end{align} 
where $A_1$ and $A_2$ are the areas of $\Sigma_{G_1}$ and
$\Sigma_{G_2}$, respectively. 
In deriving (\ref{quiver partition2}), 
we have used the assumption that $B$'s are localized at points.

Now let us evaluate (\ref{quiver partition2}) in 
the limit of $N_1\to\infty$, $N_2\to\infty$ and 
$\mu_{r+1}\to 0$ and $\nu_{s+1}\to 0$ with a suitable scaling. 
Since it is almost the
product of the partition function of gYM$_2$ on $\Sigma_{G_1}$ and that
on $\Sigma_{G_2}$, we can apply the same discussion to derive (\ref{gYM2
to Nekrasov}). 
Namely, 
we can assume that there are $2N^{(1)}_{l_1}+1$ eigenvalues 
around $q_{l_1}^{(1)}$ 
and $2N^{(2)}_{l_2}+1$ eigenvalues around $q_{l_2}^{(2)}$, respectively, 
whose distribution can be expressed as excitations from the ground states, 
{\it i.e.} the densest and symmetrical distribution of eigenvalues around
the critical points. 
When $N_1$ and $N_2$ are sufficiently large, the excitations are thought
to occur at the ``fermi levels'' at
$a^{(a)}_{l_a}=q^{(a)}_{l_a}+g_s\frac{N_{l_a}-1}{2}$ and 
$\ta^{(a)}_{l_a}=q^{(a)}_{l_a}-g_s\frac{N_{l_a}-1}{2}$ $(a=1,2)$, 
which are labeled by sets of partitions 
$(\{k_{i_1}^{l_1}\}_{l_1=1}^r, \{k_{i_2}^{l_2}\}_{l_2=1}^s)$ and 
$(\{\tk_{i_1}^{l_1}\}_{l_1=1}^r, \{\tk_{i_2}^{l_2}\}_{l_2=1}^s)$. 
In the case of $\mu_{r+1}\to0$ and $\nu_{s+1}\to0$, it is natural to
assume $N^{(a)}_{l_a}=2M_a+1$ for each $l_a$, and we can take 
the double scaling limit $M_1\to\infty$, $M_2\to\infty$ and 
$\mu_{r+1}\to 0$ and $\nu_{s+1}\to 0$ with fixing,  
\begin{align}
(g_sM_1)^{2r} \exp\left(-A_1 \mu_{r+1}(g_s M_1)^r\right) 
&\equiv \Lambda_1^{2r}, \nn \\
(g_sM_2)^{2s} \exp\left(-A_2 \mu_{s+1}(g_s M_2)^r\right) 
&\equiv \Lambda_2^{2s}.
\label{quiver double scaling limit} 
\end{align}
(See the discussion around (\ref{double scaling limit}).)
In this limit, 
the partition function (\ref{quiver partition2}) becomes 
\begin{align}
Z_{\rm quiv} = 
\left[
Z_{\rm inst}\left(\av^{(1)};\Lambda_1\right)
Z_{\rm inst}\left(\av^{(2)};\Lambda_2\right)
Z_{12}\left(\av^{(1)},\av^{(2)};M_1-M_2\right)
\right]^2, 
\end{align}
where $Z_{\rm inst}\left(\av;\Lambda\right)$ is given in 
(\ref{gYM2 to Nekrasov}) and 
\begin{align}
Z_{12}&\left(\av^{(1)},\av^{(2)};M_1-M_2\right) \nn \\
&=\sum_{\vec{\kv}^{(1)},\vec{\kv}^{(2)}}\prod_{(l_1,i_1)}\prod_{(l_2,i_2)} 
\left(a_{l_1}^{(1)}-a_{l_2}^{(2)}+g_s(M_1-M_2)
+g_s\left(k_{i_1}^{(l_1)}-k_{i_2}^{(l_2)}-i_1+i_2\right)\right)^{-1}. 
\end{align} 
Here we have ignored terms that contain $g_s(M_1+M_2)$ since it
suppresses the fluctuations by $\vec{\kv}_1$ and $\vec{\kv}_2$. 
Therefore, the partition function (\ref{quiver partition}) 
is factorized into two parts in the double scaling limit. 
Especially, under the additional condition $M_1=M_2$, 
it can be expressed as 
\begin{equation}
Z_{\rm quiv} = Z_{\rm inst}^{\rm quiv}
\left(\av^{(1)},\av^{(2)};\Lambda_1,\Lambda_2;G_1,G_2\right)^2,
\end{equation}
with 
\begin{align}
Z_{\rm inst}^{\rm quiv}
&\left(\av^{(1)},\av^{(2)};\Lambda_1,\Lambda_2;G_1,G_2\right) 
=Z_{\rm pert}^{\rm quiv}\left(\av^{(1)},\av^{(2)}\right)
\sum_{\vec{\kv}^{(1)},\vec{\kv}^{(2)}}\Lambda_1^{k_1}\Lambda_2^{k_2} \nn \\
&\times\prod_{(l_1,i_1)\neq (n_1,j_1)}
\left(\frac{a^{(1)}_{l_1}-a^{(1)}_{n_1}
+g_s(k^{(l_1)}_{i_1}-k^{(n_1)}_{j_1}+j_1-i_1)}
{a^{(1)}_l-a^{(1)}_n+g_s(j_1-i_1)}
\right)^{1-G_1} \nn \\
&\times\prod_{(l_2.i_2)\neq (n_2,j_2)}
\left(\frac{a^{(2)}_{l_2}-a^{(2)}_{n_2}
+g_s(k^{(l_2)}_{i_2}-k^{(n_2)}_{j_2}+j_2-i_2)}
{a^{(2)}_l-a^{(2)}_n+g_s(j_2-i_2)}
\right)^{1-G_2} \nn \\
&\times 
\prod_{(l_1,i_1)}\prod_{(l_2,i_2)} 
\frac
{a_{l_1}^{(1)}-a_{l_2}^{(2)}
+g_s\left(-i_1+i_2\right)}
{a_{l_1}^{(1)}-a_{l_2}^{(2)}
+g_s\left(k_{i_1}^{(l_1)}-k_{i_2}^{(l_2)}-i_1+i_2\right)}, 
\label{quiver Nekrasov formula}
\end{align}
%
and 
\begin{align}
Z_{\rm pert}^{\rm quiv}&\left(\av^{(1)},\av^{(2)}\right) \nn
 \\
&= \prod_{(l_1,i_1)\neq (n_1,j_1)}
\left(a_{l_1}^{(1)}-a_{n_1}^{(1)}+g_s(j_1-i_1)\right)^{1-G_1} 
\prod_{(l_2,i_2)\neq (n_2,j_2)}
\left(a_{l_2}^{(2)}-a_{n_2}^{(2)}+g_s(j_2-i_2)\right)^{1-G_2} \nn \\
&\hspace{1cm}\times\prod_{(l_1,i_1)}\prod_{(l_2,i_2)}
\left(a_{l_1}^{(1)}-a_{l_2}^{(2)}+g_s\left(-i_1+i_2\right)\right)^{-1},
\end{align}
where $k_a=\sum_{l_a,i_a}k_{i_a}^{(l_a)}$.  
We have again used the fact that $\ta_{l_a}^{(a)}$ can be replaced by
$a_{l_a}^{(a)}$ as long as $M_1=M_2$. 
We conjecture that the expression (\ref{quiver Nekrasov formula}) gives  
the instanton contribution to the $\cN{=}2$ $A_2$-quiver theory with the
gauge symmetry $SU(r)\times SU(s)$ with the bi-fundamental matters.

\subsection{Fundamental matters}
\label{sec:fundamental-matters}

In this subsection, we derive the instanton partition function
 for $\cN{=}2$ theories
with  the hypermultiplets in the fundamental representation from 
a two-dimensional gauge theory. 
Since we already have the formula for the quiver theory in hand, 
it can be obtained by considering a limited case of the
previous section, namely, the limit of $A_2\to\infty$. 

Let us start with the partition function (\ref{quiver partition2}). 
Before taking the double scaling limit (\ref{quiver double scaling
limit}), we perform the limit of $A_2\to\infty$. 
Then the eigenvalues $\{n_{i_2}^{(2)}\}$ cannot excite from the
ground state since the potential becomes infinitely deep effectively, 
that is, we must take 
\begin{equation}
k_{i_2}^{l_2}=0, \quad \tk_{i_2}^{l_2}=0, 
\end{equation}
for all $l_2$ and $i_{l_2}$. 
Putting
$\Lambda_1$ to $\Lambda$, $G_1$ to $G$, 
$\{a_{l_1}^{(1)}\}_{l_1=1}^r$ to $\{a_{l}\}_{l=1}^r$ and 
$\{a_{l_2}^{(2)}\}_{l_2=1}^s$ to $\{m_{f}\}_{f=1}^s$, 
the factorized partition function (\ref{quiver Nekrasov formula})
becomes 
\begin{equation}
Z_{\rm quiv}\Bigr|_{A_2\to\infty}=
Z_{\rm inst}^{\rm fund}\left(\av,\mv;\Lambda;G\right)^2
\end{equation}
with 
\begin{align}
Z_{\rm inst}^{\rm fund}
\left(\av,\mv;\Lambda;G \right) 
=Z_{\rm pert}^{\rm fund}\left(\av,\mv\right)
\sum_{\vec{\kv}}\Lambda^{k} 
&\prod_{(l,i)\neq (n,j)}
\left(\frac{a_{l}-a_{n}
+g_s(k^{(l)}_{i}-k^{(n)}_{j}+j-i)}
{a_l-a_n+g_s(j-i)}
\right)^{1-G} \nn \\
&\times 
\prod_{(l,i)}\prod_{(f,j)} 
\frac
{a_{l}-m_f
+g_s\left(-i+j\right)}
{a_{l}-m_f
+g_s\left(k_{i}^{(l)}-i+j\right)}, 
\label{fundamental Nekrasov formula}
\end{align}
where
\begin{align}
Z_{\rm pert}^{\rm fund}\left(\av,\mv\right)
&= \prod_{(l,i)\neq (n,j)}
\left(a_{l}-a_{n}+g_s(j-i)\right)^{1-G} 
\prod_{(l,i)}\prod_{(f,j)}
\left(a_{l}-m_f+g_s\left(-i+j\right)\right)^{-1}. 
\end{align}
If we set $G=0$, (\ref{fundamental Nekrasov formula}) is nothing but the
instanton counting for $\cN{=}2$ theory with massive fundamental
hypermultiplets given in \cite{Nekrasov:2003rj}. 
This fact supports our conjecture presented in the previous subsection.

\section{Continuum Limit}
\label{sec:relation-c=0-matrix}

\subsection{Eigenvalue density vs profile function}

We first would like to mention on the relationship between the
eigenvalue density and the profile function introduced in
\cite{Nekrasov:2003rj}. Both quantities are essentially connected
through difference equations. The profile function expression of the
discrete matrix model is useful to see a series expansion with respect
to $g_s$, which gives the graviphoton corrections to $\N{=}2$ SUSY gauge
theory.

We now define an eigenvalue density for (\ref{discrete MM})
in the standard way, 
\be
\rho(x) \equiv \frac{1}{N}\sum_{i=1}^{N}\delta(x-g_s n_i), 
\label{eigenvalue density}
\ee
which follows $\int dx \rho(x) =1$.
The Vandermonde determinant part
can be written in terms of the eigenvalue density as 
\be
\prod_{1 \leq i<j \leq N}(g_s n_i- g_s n_j)^{2-2G}
=\exp\left\{
(1-G)N^2\pint dx dy \log(x-y)\rho(x) \rho(y)
\right\},
\label{Vandermonde integral}
\ee
where the integral stands for a principal part.
Similarly, the potential part is
\be
\Tr W(\Phi) = N\int dx W(x) \rho(x).
\ee
Thus we can write the partition function of the discrete matrix model
(\ref{discrete MM}) as
\begin{multline}
Z=\sum_{\{\rho(x)\}}
\exp\Big\{
(1-G) N^2 \pint dx dy \log(x-y)\rho(x)\rho(y)
-\frac{A N}{g_s}\int dx W(x)\rho(x) 
\Big\}, 
\label{discrete MM integral}
\end{multline}
where $\{\rho(x)\}$ denotes the possible sets of the eigenvalue density 
$\rho(x)$. 

Here we introduce a function $\gamma_{g_s}(x)$ which obeys the
following difference equation with respect to the finite parameter
$g_s$; 
\be
\gamma_{g_s}(x+g_s)+\gamma_{g_s}(x-g_s)-2\gamma_{g_s}(x)
=\log x. 
\label{gamma}
\ee
Then the integral in (\ref{Vandermonde integral}) can be rewritten as 
\begin{align}
\pint dx dy &\log(x-y) \rho(x) \rho(y) \nn \\
&=\pint dx dy
\Bigl(\gamma_{g_s}(x-y+g_s)+\gamma_{g_s}(x-y-g_s)-2\gamma_{g_s}(x-y)\Bigr)
\rho(x) \rho(y) \nn \\
&=-\pint dx dy\,
\gamma_{g_s}(x-y)\drho(x) \drho(y),
\end{align}
where 
\begin{equation}
\drho(x)\equiv \rho(x+\frac{g_s}{2})-\rho(x-\frac{g_s}{2}), 
\label{delta rho}
\end{equation}
and
we have used the fact that the integral measure is invariant
under a constant shift $x\rightarrow x\pm g_s/2$.
Similarly, the potential term can be expressed as 
\be
\Tr W(\Phi) = -N\int dx\, U(x) \drho(x),
\ee
using the function $U(x)$ satisfying the relation, 
\be
U(x+\frac{g_s}{2})-U(x-\frac{g_s}{2}) = W(x).
\label{U}
\ee
For $r$-th order polynomial potential, we generally have
the $(r+1)$-th order polynomial $U(x)$. 
For example, 
for the quadratic potential $W(x)=\frac{\mu}{2}x^2$, 
$U(x)$ is given by 
\be
U(x) = \frac{\mu}{2}
\left(
\frac{1}{3g_s}x^3-\frac{g_s}{12}x
\right),
\ee
up to an integral constant. 
Then the partition function (\ref{discrete MM integral}) 
is expressed as 
\begin{align}
Z=\sum_{\{\rho(x)\}} \exp\Bigl\{
&-(1-G)N^2\pint dx dy \,
\gamma_{g_s}(x-y)\drho(x) \drho(y)
+\frac{A N}{g_s} \int dx U(x)\drho(x)
\Bigr\}, 
\label{discrete MM in profile}
\end{align}
using the expression $\drho(x)$.

By the way, there is another useful instrument to describe 
the discrete matrix model; the profile function of the 
Young diagram utilized in \cite{Nekrasov:2003rj},%
\begin{align}
f_{\kv}(x;g_s)&\equiv
|x|+\sum_{i=1}^{\infty}\Bigl[
|x-g_s(k_i-i+1)|-|x-g_s (k_i-i)|+|x+g_s i|-|x+g_s(i-1)|
\Bigr] \nn \\
&= \sum_{i=1}^{\infty}\Bigl[
\left|x-g_s(k_i-i+1)\right|-\left|x-g_s(k_i-i)\right|
\Bigr],
\label{profile function}
\end{align}
where $\{k_i\}$ is a sequence of non-increasing 
non-negative integers. 
We assume that $k_i=0$ for $i>N$. 
This function is closely related to $\drho(x)$ introduced above. 
To see it, let us consider the case where the discrete eigenvalues 
$\{n_i\}$ are decomposed into $r$ decoupled lumps starting from 
$a_l$ $(l=1,\cdots,r)$.   
The corresponding eigenvalue density is 
\begin{align}
\rho(x)=\frac{1}{N}\sum_{l=1}^{r}\sum_{i_l=1}^{N_l}
\delta\left(x-a_l-g_s\left(k_{i_l}^{(l)}-i+\frac{1}{2}\right)\right), 
\label{colored eigenvalue density}
\end{align}
where $N=N_1+\cdots+N_r$, 
$\{k_{i_l}^{(l)}\}_{i_l=1}^{N_l}$ 
is a sequence of non-increasing non-negative integer, 
and $1/2$ in the delta function is a convention. 
Particularly, if we define the eigenvalue density for 
the ``ground state'' by setting all $k^{(l)}_i$ to be zero,
\be
\rho_0(x)=\frac{1}{N}\sum_{l=1}^{r}\sum_{i_l=1}^{N_l}
\delta\left(x-a_l+g_s\left(i_l-\frac{1}{2}\right)\right), 
\ee
we find
\begin{align}
\drho_0(x) &\equiv 
\rho_0(x+\frac{g_s}{2})-\rho_0(x-\frac{g_s}{2}) \nn \\
&= \frac{1}{N}\sum_{l=1}^{r}\Bigl(
\delta(x-a_l+g_sN_l)-\delta(x-a_l)
\Bigr), 
\label{rho zero}
\end{align}
since the intermediate $\delta$-functions are canceled with each other. 
Using (\ref{colored eigenvalue density}), 
we can easily show that $\drho(x)$ and 
the (colored) profile function is related by
\begin{align}
-\frac{1}{2N}\sum_{l=1}^{r}f''_{\kv^{(l)}}(x-a_l) 
&= \drho(x) -\frac{1}{N}\sum_{l=1}^{r}\delta(x-a_l+g_sN_l),  
\label{relation between profile and ev}
\end{align}
where the second term in r.h.s. is a counterpart of the contribution
to the summation from $N_l$ to infinity.

Now we have a dictionary to translate the eigenvalue density 
into the profile function in hand. 
As an application, let us write down the normalized 
Vandermonde determinant (measure part), 
\begin{equation}
\mu_{\vec{\kv}}^{2}(\av,g_s) \equiv
\prod_{(l,i_l)\neq(n,j_n)}
\left(\frac{
a_l-a_n+g_s(k^{(l)}_{i_l}-k^{(n)}_{j_n}+j_n-i_l)
}{
a_l-a_n+g_s(j_n-i_l)}
\right),  
\label{measure part}
\end{equation}
using both of the eigenvalue density and the profile function. 
In terms of the eigenvalue density, (\ref{measure part}) can be written
as 
\begin{equation}
\mu_{\vec{\kv}}^{2}(\av,g_s) =\exp\left\{
-N^2\pint dx dy\,
\gamma_{g_s}(x-y)(\drho(x) \drho(y)-\drho_0(x) \drho_0(y))
\right\}. 
\end{equation}
On the other hand, it can be also written as 
\begin{align}
\mu_{\vec{\kv}}^{2}(\av,g_s)
=\exp\Bigg\{
&-\frac{1}{4}\pint dx dy
\gamma_{g_s}(x-y)\sum_{l \neq n}
f''_{\kv^{(l)}}(x-a_l)f''_{\kv^{(n)}}(x-a_n) \nn \\
&+\int dx \sum_{l,n}
\gamma_{g_s}(x-a_l+g_sN_l)\left(f''_{\kv^{(n)}}(x-a_n)
-2\delta(x-a_n)\right) \nn \\
&+\sum_{l \neq n}\gamma_{g_s}(a_l-a_n)
\Bigg\}, 
\end{align}
using the profile function. The second term in the exponential
represents the finite $N$ effect (cut-off at $g_s N_l$) and the third is
the perturbative contribution to the prepotential. These results also agree
with \cite{Matsuo:2004cq} up to irrelevant constants.

\subsection{Dijkgraaf-Vafa theory and Douglas-Kazakov phase transition}

The discrete version of the matrix model (\ref{discrete MM})
has obviously a continuous limit of $g_s\rightarrow 0$.
In this limit, the summation over a set of possible integers
reduces to an integration over continuous variables
if we restrict the case as $G=0$ (on sphere), 
\be
Z=\int \prod_{i=1}^{N}d\lambda_i
\prod_{i<j}(\lambda_i-\lambda_j)^2
e^{-\frac{1}{{\tg}_s}\sum_i W(\lambda_i)},
\label{continuous MM}
\ee
where $\lambda_i\equiv g_s n_i$ and
we have rescaled the coupling constant as
$g_s/A = \tg_s$.
So we have an ordinary type of the Hermitian (holomorphic)
matrix model with the potential (action) $W$.
Following \cite{Dijkgraaf:2002fc,Dijkgraaf:2002vw,Dijkgraaf:2002dh},
we can obtain an $\N{=}1$ effective superpotential
from  a free energy of the matrix model (\ref{continuous MM})
in a large $N$ limit by
\be
W_{\rm eff} = \sum_i N_i \frac{\del {\cal F}_0(S_i)}{\del S_i}
+ 2\pi i \tau \sum_i S_i,
\ee
where ${\cal F}_0$ is a planar contribution to the free energy of
the matrix model and $S_i \equiv \tg_s \hat{N}_i$ are fixed for
large matrix sizes $\hat{N}_i$ and identified with glueball
superfields $S_i =\Tr {\cal W}^{(i)}_\alpha {\cal W}^{(i)\alpha}$
for $i$-th $U(1)$ sector.

This continuum limit can be understood from a point of view of the
profile function mentioned in the previous subsection. All difference
equations with the spacing $g_s$ turn to differential equations. 
For example, eqs.~(\ref{gamma}), (\ref{delta rho}) and (\ref{U}) become
\begin{align}
g_s^2\gamma_{g_s}''(x)&\to \log x,
\label{gamma''}\\
g_s^{-1}\drho(x)&\to \rho'(x),\\
g_s U'(x) & \to W(x),
\label{U'}
\end{align}
in the continuum limit $g_s\rightarrow 0$, respectively.
Thus the partition function of the discrete matrix model in terms of the
profile function (\ref{discrete MM in profile}) reduces to
\begin{multline}
Z\to\int {\cal D}\rho\exp\left\{
(1-G)N^2\pint dx dy \log(x-y)\rho(x)\rho(y)
-\frac{N}{\tg_s}\int dx W(x) \rho(x)
\right\},
\end{multline}
where we use the relations (\ref{gamma''})-(\ref{U'})
and a partial integral from the first line to the second line.

Similarly, the partition functions including various type of matter
discussed in the previous section
agree with the Dijkgraaf-Vafa theory with matters in the continuum
limit, for the adjoint matter:
\bea
Z_{\rm adj}&=&\int d\Phi dX dY e^{-\frac{1}{\tg_s}
\Tr\left[
\Phi[X,Y]+m XY+W(\Phi)
\right]}\nn\\
&=&\int \prod_i d\lambda_i
\frac{\prod_{i\neq j}(\lambda_i-\lambda_j)}
{\prod_{i,j}(m+\lambda_i-\lambda_j)}
 e^{-\frac{1}{\tg_s}
\sum_i W(\lambda_i)
},
\eea
the $A_2$ quiver theory:
\begin{align}
Z_{\rm quiver}&=\int d\Phi_1 d\Phi_2 dB d\tilde{B}
e^{-\frac{1}{\tg_s}
\Tr\left[
\tilde{B}\Phi_1 B - B \Phi_2 \tilde{B}
+W_1(\Phi_1)+W_2(\Phi_2)
\right]}\nn\\
&=\int \prod_i d\lambda_i \prod_i d\nu_i
\frac{\prod_{i\neq j}(\lambda_i-\lambda_j)
\prod_{i\neq j}(\nu_i-\nu_j)}{\prod_{i,j}(\lambda_i-\nu_j)}
e^{-\frac{1}{\tg_s}\left[
\sum_i W_1(\lambda_i) + \sum_i W_2(\nu_i)
\right]}, 
\end{align}
and the fundamental matters:
\begin{align}
Z_{\rm fund}&=\int d\Phi dB d\tilde{B}
e^{-\frac{1}{\tg_s}
\Tr\left[
\tilde{B}\Phi B + m\tilde{B}B + W(\Phi)\right]}\nn\\
&=\int \prod_i d\lambda_i
\frac{\prod_{i \neq j}(\lambda_i-\lambda_j)}
{\prod_i\prod_f(m_f+\lambda_i)}
e^{-\frac{1}{\tg_s}\sum_i W(\lambda_i)}. 
\end{align}
Thus it seems that we can reproduce all results 
of the effective superpotential for
$\N{=}1$ theory from the continuum limit of the discrete matrix model,
which also produces the prepotential for $\N{=}2$ theory with the same
matter contents.

However these continuum limits are too naive since the discrete matrix
model has an unavoidable constraint, 
\be
n_i-n_{i+1}\geq 1,
\label{constraint}
\ee
for any $i$.
Following \cite{Douglas:1993ii}, let us consider
the continuum limit $g_s\rightarrow 0$ and large
$N$ limit simultaneously, with fixing the following quantities:
\be
t \equiv g_s i, \qquad n(t)\equiv g_s n_i.
\ee
The parameter $t$ has a range of $0\leq t \leq S\equiv g_s N$.
The constraint (\ref{constraint}) means
\be
n'(t) \geq -1,
\ee
in this continuum limit and it gives the
constraint for the eigenvalue density:
\be
\rho(x)\leq 1,
\ee
since the $n(t)$ and $\rho(x)$ are related by
\be
\rho(x)=-\frac{dn^{-1}(x)}{dx}=-\frac{1}{n'(t)}.
\ee
Therefore we must carefully take account of the constraint
for the eigenvalue density when we solve the discrete matrix model
in the continuum and large $N$ limit.

This essential difference
between the discrete matrix model and ordinary continuous matrix model
causes the third order phase transition, which is called as
the Douglas-Kazakov phase transition \cite{Douglas:1993ii}.
This phase transition occurs when maximum value of the eigenvalue density
reaches at the upper limit $1$. The Wigner's semi-circle type solution
does not admit in this phase.

On the other hand, there exists another type of 
third order phase transition in 
the ordinary continuous matrix models, namely, 
the Gross-Witten phase transition.  
Originally, this type of
the phase transition was found in a unitary matrix model, which is
the dimensionally reduced matrix model of two-dimensional 
Yang-Mills theory on the lattice. 
It occurs when 
the end points of the lump of the eigenvalue density meet at the
opposite side on a circle since the eigenvalues distribute 
on the circle due to the unitarization. 
If we consider a covering space of the unitary circle, 
we can regard the phase transition as the joining of 
the cuts on a line of the covering space.
In this sense, we can say that the Gross-Witten phase transition occurs 
when the end points of lumps of eigenvalues joins with each other 
in the continuous matrix models.

\begin{figure}[ht]
\begin{center}
\begin{tabular}{ccc}
\includegraphics[scale=0.35]{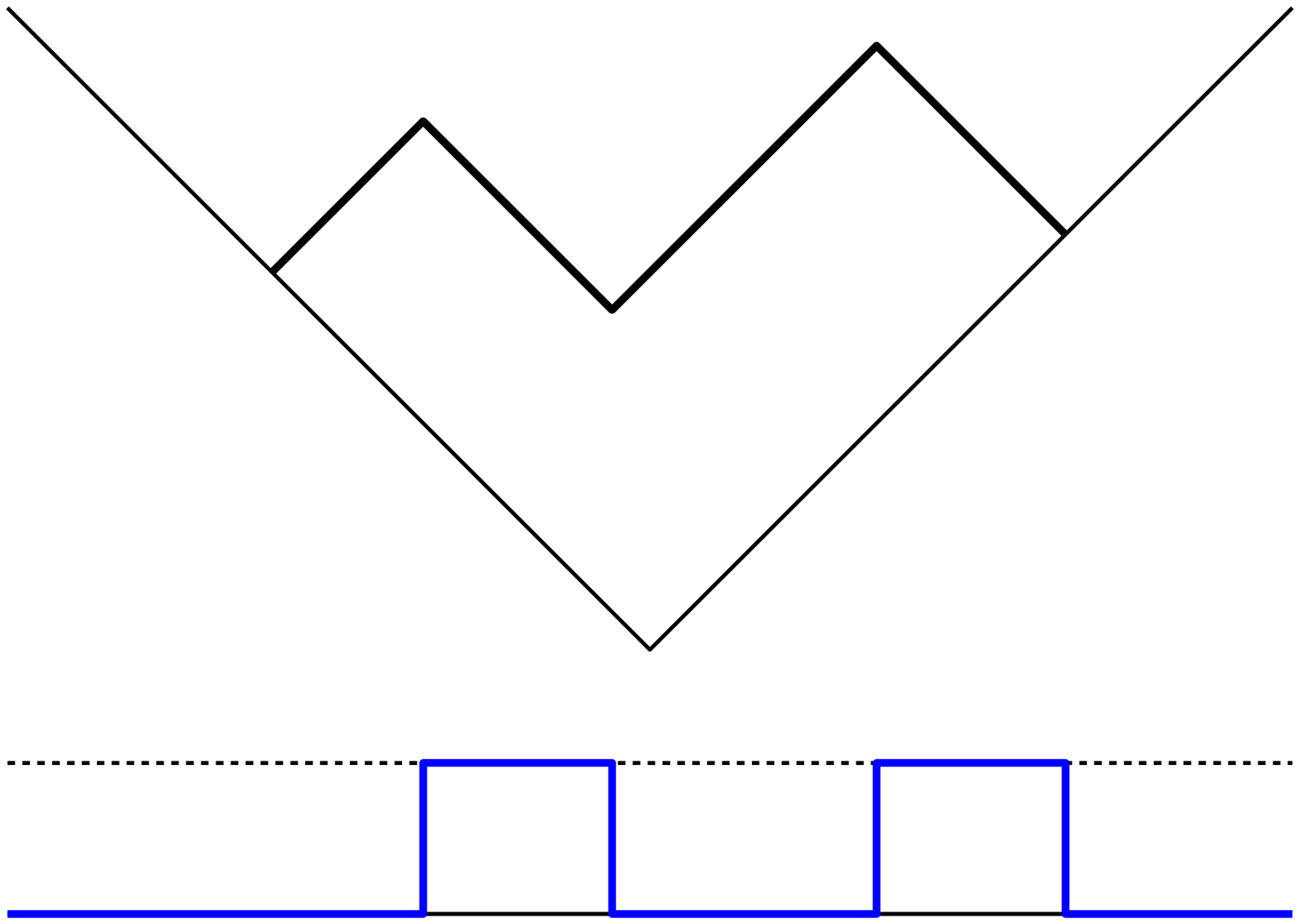}&\hspace{1cm}&
\includegraphics[scale=0.35]{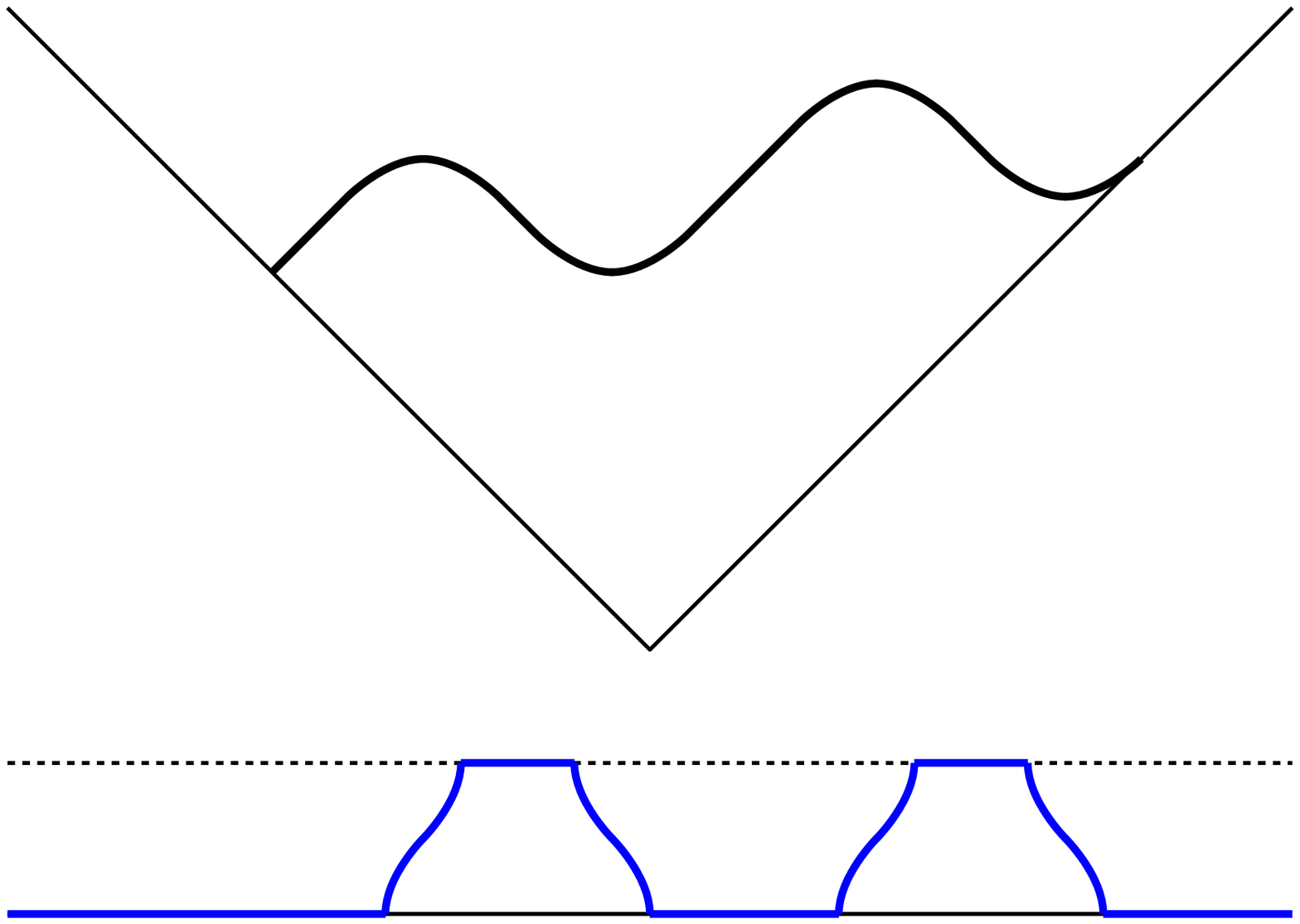}\\
(a) Ground state & &(b) Douglas-Kazakov phase   \\
&&\\
\includegraphics[scale=0.35]{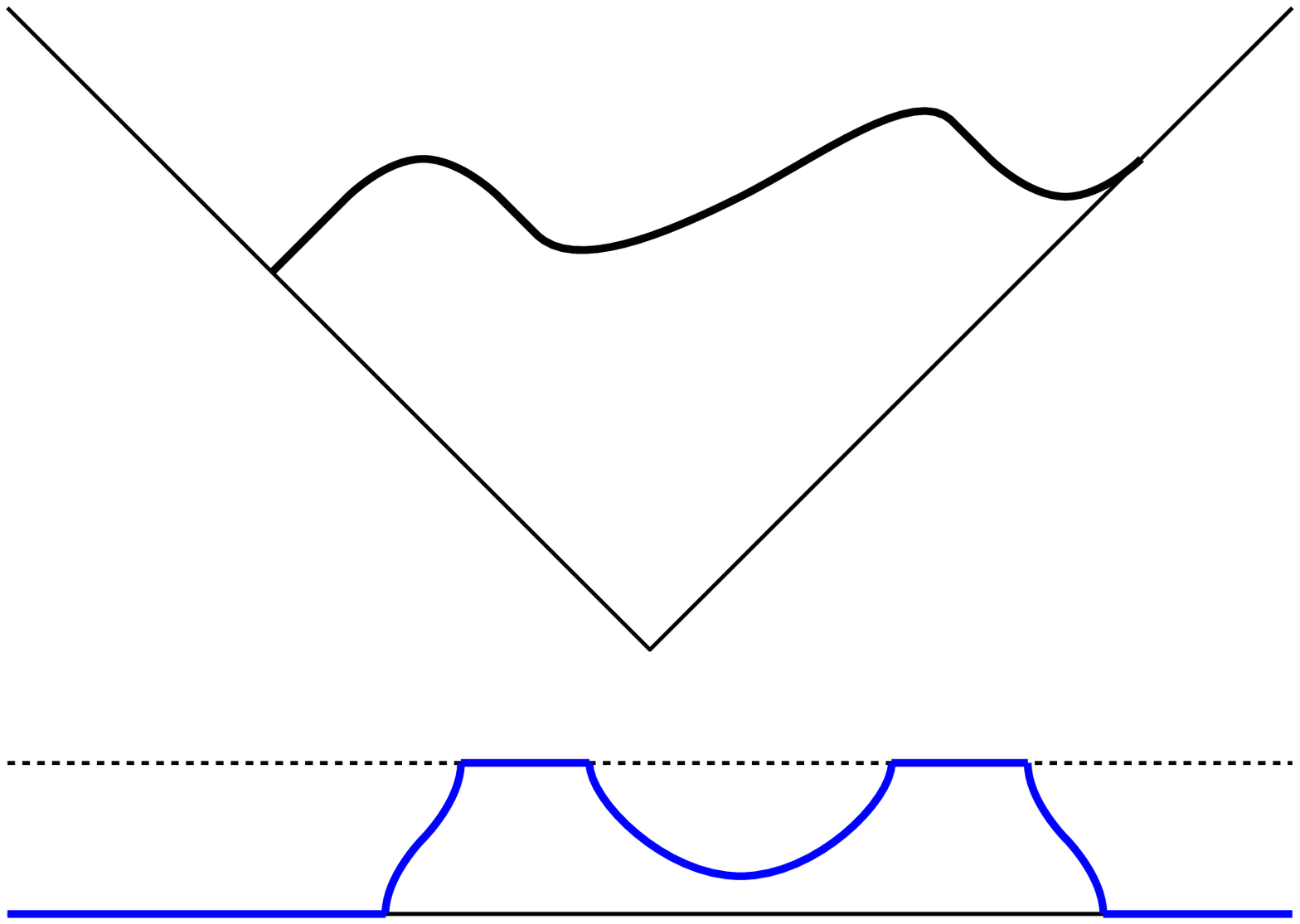}&\hspace{1cm}&
\includegraphics[scale=0.35]{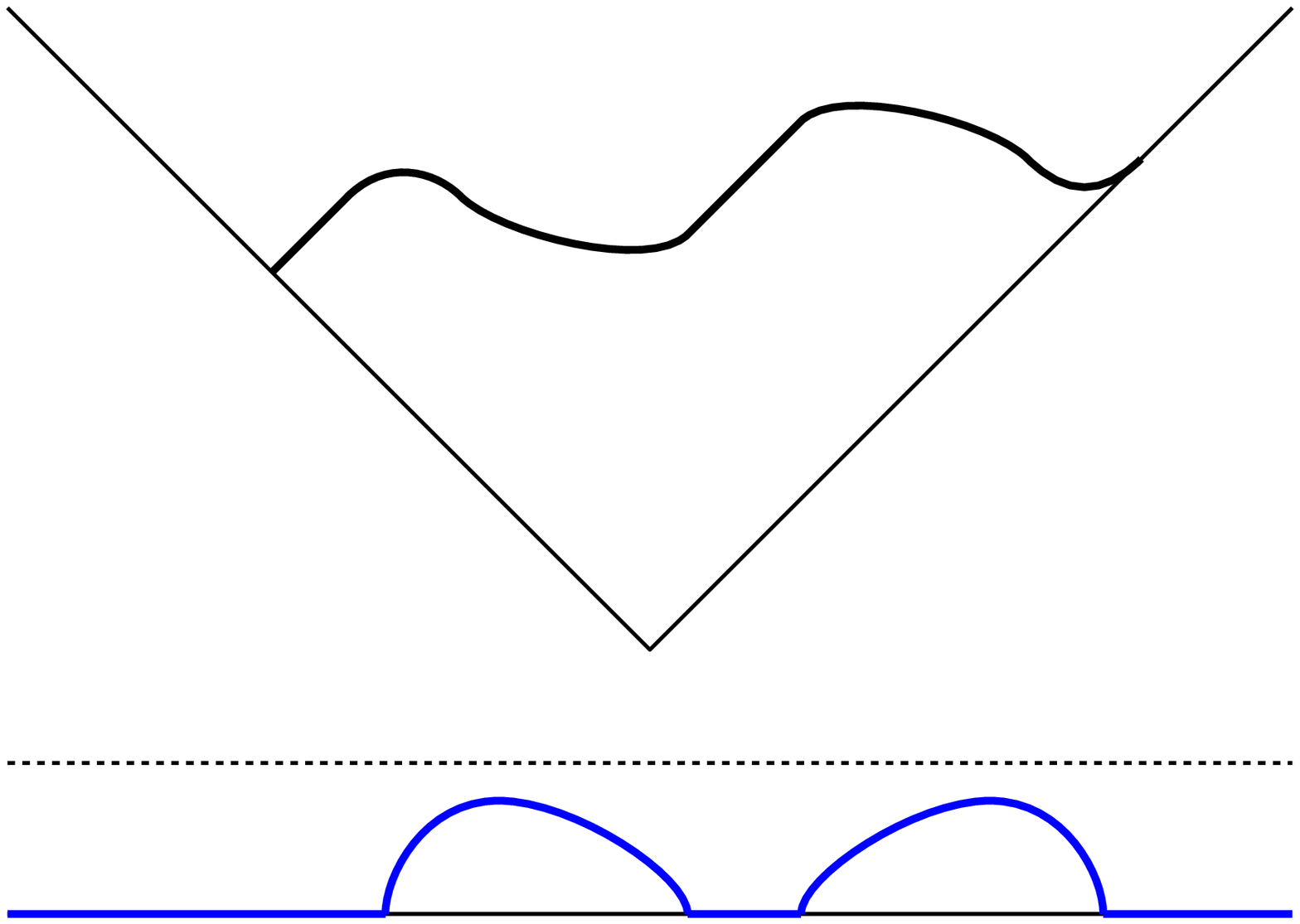}\\
(c) Gross-Witten phase & &(d) Matrix model phase
\end{tabular}
\end{center}
\caption{Various phases in the continuum limit of the discrete matrix
 model. Here we show a two-cuts solution for instance.
 }
\label{Various phases}
\end{figure}

In Fig.\ref{Various phases}, we sketch 
the shapes of the Young diagrams (the upper figures) 
and the corresponding eigenvalue densities (the lower figures)
in various phases. 
Looking at these figures, 
we notice that the shape of each Young diagram is constructed from 
three pieces: 
(1) a right-down planar edge with 45 degree, 
(2) a right-up planar edge with 45 degree, 
and (3) a melting corner. 
Correspondingly, the shape of each eigenvalue density is 
constructed from 
(1) a ``ceiling'' (a flat region with the upper limit), 
(2) a ``floor'' (a flat region with $\rho=0$), 
and (3) a curve connecting them. 
In fact, a right-down planar edge of the Young diagram 
corresponds to the densest eigenvalues on every $g_s$ interval, 
and the right-up planar edge of the Young diagram corresponds to the
complete vacant space (no eigenvalues region), 
as typically like as the ground state distribution (a). 
As the corner of a Young diagram is melting 
(eigenvalues of the free fermions are exciting from the fermi surface), 
the planar edge regions get narrower. 
At the end, 
when at least one of the planar edge of the Young diagram disappears, 
a third order phase transition occurs as we mentioned above. 
The disappearance of a right-down edge means that 
the ceilings of the eigenvalue density are pinched and 
the Douglas-Kazakov phase transition occurs. 
Similarly, if a right-up planar edge disappears,  
the Gross-Witten phase transition occurs. 

From these observations, we notice that the following important fact: 
If we exchange a role of the right-down and right-up edge 
in the Young diagram, 
or correspondingly, 
the positions of eigenvalues (electron) and vacuums (hall),
the Douglas-Kazakov phase transition and 
the Gross-Witten phase transition exchange each other.%
\footnote{
For another discussion on the duality between the Douglas-Kazakov phase
transition and the Gross-Witten phase transition, see
\cite{Gross:1994ub}. 
} 
In this sense, these phase transitions are essentially 
the same phenomena and are dual under the above operation.

Finally, we comment on the relation to the Nekrasov's instanton 
partition function and Dijkgraaf-Vafa theory. 
As we discussed in section 3, we need to take
a suitable double scaling large $N$ limit in order to obtain the instanton
partition function from the discrete matrix model ($\gYM$). In this
limit, the chiral decomposition is also needed. To decompose the
partition function into two parts, we assume the melting of the corner of
the Young diagram is sufficiently small comparing with the planar
right-down edge. We have scaled up the planar region and decoupled the
two fermi surface with each other. Namely, we need to take the large $N$
limit at least in the phase (b). The scaled-up melting corner of one of
the fermi surface will correspond to the limiting shape of the instanton
partition function.
On the other hand, the holomorphic matrix model using in 
the Dijkgraaf-Vafa theory 
does not have the Douglas-Kazakov type phase transition since
there is no upper limitation in the eigenvalue densities. 
So all eigenvalue densities have the semi-circle shape. 
Thus, we find the Dijkgraaf-Vafa analysis is valid only for the phase (d). 
We need to take the large $N$ and continuum limit in this phase 
to get the effective superpotential of $\N{=}1$ theories.

We can obtain all order graviphoton corrections
for the ${\cal N}{=}2$ prepotential
from an asymptotic expansion of Nekrasov's formula.
It however is still difficult to obtain all order graviphoton correction
for the superpotential of ${\cal N}{=}1$
theory since we need sub-leading terms in the $1/N$ expansion of the
random matrix models, where we can not use the WKB
(saddle point) approximation.
The discretization of the random matrix model seems
to connect the two approaches by Nekrasov and Dijkgraaf-Vafa. 
We expect that our observation sheds new lights
on the exact relation between them and gives a new closed formula
for the effective superpotential of ${\cal N}{=}1$ theory
which gives all graviphoton corrections
as the asymptotic expansion like Nekrasov's formula,
beyond the correspondence between $\cN{=}2$ and $\cN{=}1$ theories
at the algebraic geometry (curve) level \cite{Cachazo:2002pr}.

\section*{Acknowledgements}
The authors would like to thank H.~Kawai, T.~Tada, T.~Matsuo, 
T.~Kuroki, Y.~Shibusa, Y.~Tachikawa, H.~Fuji and H.~Kanno
for useful discussions and valuable comments. 
This work is supported by Special Postdoctoral Researchers
Program at RIKEN.

\appendix

\section{Generalized $\zeta$-function and perturbative contribution}

The perturbative contribution to the partition function
is obtained essentially from infinite products, 
\be
\prod_{(i,j)\in \IN^2}(a_l-a_n+g_s(j-i)),
\label{pert}
\ee
where $\IN$ is a set of natural numbers (non-zero positive integers).
This type of the infinite product can be understood
as the determinant of an operator (the product of eigenvalue spectra).
This infinite product can be evaluated by using the $\zeta$-function
regularization technique.

We define the generalized $\zeta$-function as 
\be
\zeta_{\vec{\e}}(s;x) \equiv \sum_{\vec{n}\in \IN^d}
\frac{1}{(x+\vec{\e}\cdot\vec{n})^s},
\ee
where $\e=(\e_1,\e_2,\ldots,\e_d)$ is a vector of parameters and
$n=(n_1,n_2,\ldots,n_d)$ is a set of non-zero positive integers.
Using this $\zeta$-function, the logarithm of infinite products
like (\ref{pert}) can be expressed as
\bea
\log \prod_{\vec{n}\in \IN^d}(x+\vec{\e}\cdot\vec{n})
&=& \sum_{\vec{n}\in \IN^d} \log (x+\vec{\e}\cdot\vec{n})\nn\\
&=& -\left.\frac{d}{ds}\zeta_{\vec{\e}}(s;x)\right|_{s\rightarrow 0}.
\eea
Multiplying the gamma function,
\be
\Gamma(s) = \int_{0}^{\infty}u^{s-1}e^{-u} du,
\ee
by $\zeta_{\vec{\e}}(s;x)$, we find
\be
\Gamma(s)\zeta_{\vec{\e}}(s;x)
=\sum_{\vec{n}\in \Z_{+}^d} \int_{0}^{\infty} du
 (x+\vec{\e}\cdot\vec{n})^{-s}
u^{s-1}e^{-u}.
\ee
After changing variable by $u=t(x+\vec{\e}\cdot\vec{n})$,
we obtain
\bea
\zeta_{\vec{\e}}(s;x) &=& \frac{1}{\Gamma(s)}
\int_{0}^{\infty} \frac{dt}{t} t^s
\sum_{\vec{n}\in \Z_{+}^d}e^{-(x+\vec{\e}\cdot\vec{n})t}\nn\\
&=& \frac{1}{\Gamma(s)}
\int_{0}^{\infty} \frac{dt}{t} t^s
\frac{e^{-tx}}{\prod_{i=1}^{d}(e^{\e_i t}-1)}.
\eea

Especially, in the case of (\ref{pert}), we have
\be
\prod_{(i,j)\in \IN^2}(a_l-a_n+g_s(j-i))
=e^{-\gamma_{g_s}(a_l-a_n)},
\ee
where
\be
\gamma_{g_s}(x)=
-\left.\frac{d}{ds}
\left(
\frac{1}{\Gamma(s)}
\int_{0}^{\infty} \frac{dt}{t} t^s
\frac{e^{-tx}}{(e^{g_s t}-1)(e^{-g_s t}-1)}
\right)
\right|_{s\rightarrow 0}.
\ee
It is easy to find from this expression to show  
that it satisfies the difference equation, 
\be
\gamma_{g_s}(x+g_s)+\gamma_{g_s}(x-g_s)-2\gamma_{g_s}(x)
=\log x,
\ee
and that the series expansion with respect to $g_s$ becomes
\be
\gamma_{g_s}(x) = \sum_g g_s^{2g-2} f_g(x),
\ee
with
\be
\begin{split}
& f_0(x) = \frac{1}{2}x^2 \log x -\frac{3}{4}x^2,\\
& f_1(x) = -\frac{1}{12}\log x,\\
& f_2(x) = -\frac{1}{240}\frac{1}{x^2},\\
& \qquad \vdots\\
& f_g(x) = \frac{B_{2g}}{2g(2g-2)x^{2g-2}}.
\end{split}
\ee

\bibliographystyle{JHEP}
\bibliography{refs}

\end{document}